\documentclass[a4paper,11pt]{article}
\usepackage{jheppub}
\pdfoutput=1
\usepackage{amsmath,amssymb,amsfonts}
\usepackage{hyperref}

\newcommand{\nc}{\newcommand}
\nc{\tred}[1]{\textcolor{red}{#1}}
\nc{\tblue}[1]{\textcolor{blue}{#1}}
\nc{\tgreen}[1]{\textcolor{green}{#1}}
\nc{\tpurple}[1]{\textcolor{purple}{#1}}
\nc{\btred}[1]{\textcolor{red}{\bf #1}}
\nc{\btblue}[1]{\textcolor{blue}{\bf #1}}
\nc{\btgreen}[1]{\textcolor{green}{\bf #1}}
\nc{\btpurple}[1]{\textcolor{purple}{\bf #1}}
\nc{\ncsha}{{\mbox{\cyr X}^{\mathrm NC}}} \nc{\ncshao}{{\mbox{\cyr	X}^{\mathrm NC}_0}}

\newcommand{\delete}[1]{}

\delete{% Use the next lines to suppress names
	\nc{\mlabel}[1]{\label{#1}}
	\nc{\mcite}[1]{\cite{#1}}
	\nc{\mref}[1]{\ref{#1}}
	\nc{\meqref}[1]{\eqref{#1}}
	\nc{\mbibitem}[1]{\bibitem{#1}}
}

	\nc{\mlabel}[1]{\label{#1}{\hfill \hspace{1cm}{\bf{{\ }\hfill(#1)}}}}
	\nc{\mcite}[1]{\cite{#1}{{\bf{{\ }(#1)}}}}
	\nc{\mref}[1]{\ref{#1}{{\bf{{\ }(#1)}}}}
	\nc{\meqref}[1]{\eqref{#1}{{\bf{{\ }(#1)}}}}
	\nc{\mbibitem}[1]{\bibitem[\bf #1]{#1}}
\nc{\ot}{\otimes}
\nc{\name}[1]{{\bf #1}}
\nc{\im}{\mathrm{im}}

\nc{\sha}{{\mbox{\cyr X}}}  %used to be \cyr
\newfont{\scyr}{wncyr10 scaled 550}
\nc{\ssha}{\mbox{\bf \scyr X}}
\nc{\shap}{{\mbox{\cyrs X}}} %sha as product
\nc{\shpr}{\diamond}    %Shuffle product
\nc{\shp}{\ast} \nc{\shplus}{\shpr^+}
\nc{\shprc}{\shpr_c}    %Cartier's product
\nc{\dep}{\mrm{dep}} \nc{\lc}{\lfloor} \nc{\rc}{\rfloor}
\nc{\db}{\leq_{\rm db}} \nc{\bfk}{\bf k}

\nc{\Id}{\mathrm{Id}}

\nc{\Aut}{\mathrm{Aut}}
\nc{\Der}{\mathrm{Der}}

\nc{\CC}{\mathbb{C}}
\nc{\NN}{\mathbb{N}}
\nc{\QQ}{\mathbb{Q}}
\nc{\RR}{\mathbb{R}}
\nc{\ZZ}{\mathbb{Z}}

\nc{\cala}{{\mathcal A}} \nc{\calb}{{\mathcal B}}
\nc{\calc}{{\mathcal C}}
\nc{\cald}{{\mathcal D}} \nc{\cale}{{\mathcal E}}
\nc{\calf}{{\mathcal F}} \nc{\calg}{\mathcal{G}}
\nc{\calh}{{\mathcal H}} \nc{\cali}{{\mathcal I}}
\nc{\call}{{\mathcal L}} \nc{\calm}{{\mathcal M}}
\nc{\caln}{{\mathcal N}} \nc{\calo}{{\mathcal O}}
\nc{\calp}{{\mathcal P}} \nc{\calr}{{\mathcal R}}
\nc{\cals}{{\mathcal S}} \nc{\calt}{{\mathcal T}}
\nc{\calu}{{\mathcal U}} \nc{\calw}{{\mathcal W}} \nc{\calk}{{\mathcal K}} 
\nc{\calx}{{\mathcal X}} \nc{\CA}{\mathcal{A}}

\nc{\fraka}{{\mathfrak a}} \nc{\frakA}{{\mathfrak A}}
\nc{\frakb}{{\mathfrak b}} \nc{\frakB}{{\mathfrak B}}
\nc{\frakc}{{\mathfrak c}}
\nc{\frakD}{{\mathfrak D}} \nc{\frakF}{\mathfrak{F}}
\nc{\frakf}{{\mathfrak f}} \nc{\frakg}{{\mathfrak g}}
\nc{\frakH}{{\mathfrak H}} \nc{\frakL}{{\mathfrak L}}
\nc{\frakM}{{\mathfrak M}} \nc{\bfrakM}{\overline{\frakM}}
\nc{\frakm}{{\mathfrak m}} \nc{\frakP}{{\mathfrak P}}
\nc{\frakN}{{\mathfrak N}} \nc{\frakp}{{\mathfrak p}}
\nc{\frakS}{{\mathfrak S}} \nc{\frakT}{\mathfrak{T}}
\nc{\frakX}{{\mathfrak X}}
\nc{\frakZ}{\mathfrak{Z}}
\nc{\frakJ}{\mathfrak{J}}
\nc{\frakR}{\mathfrak{R}}
\nc{\GL}{\mathrm{GL}}
\nc{\gl}{\mathfrak{gl}}
\nc{\frakh}{\mathfrak{h}}

\font\cyr=wncyr10 \font\cyrs=wncyr7

\nc{\cocont}[1]{_{\prec #1}}

\nc{\lin}{L}
\nc{\mlin}{\mathbf{L}}
\nc{\nsha}{\diamond}
\nc{\oF}{\overline{F}}

\usepackage[svgnames,dvipsnames]{xcolor}
\definecolor{labelcolor}{RGB}{194, 175, 116}
\definecolor{rmkcolor}{RGB}{15,120,255}

\usepackage[final]{showlabels} % use "inline" or "final"

\newif\ifToggleMacros
\ToggleMacrostrue  %% View comments
\ifToggleMacros
    \newcommand{\lir}[1]{{\color[RGB]{240,10,8} {LG:#1}}}
    \newcommand{\jwb}[1]{{\color[RGB]{10,15,245} {JW:#1}}}
    \newcommand{\jianrong}[1]{{\color[RGB]{0,180,20} {Jianrong:#1}}}
    \newcommand{\jh}[1]{{\color[RGB]{0,142,236} {JH:#1}}}
    \newcommand{\sooo}[1]{{\color[RGB]{0,128,128} {Sungsoo:#1}}}
    \newcommand{\sml}[1]{{\color[RGB]{238,130,238} {Sangmin:#1}}}
\else
   \newcommand{\lir}[1]{}
   \newcommand{\jwb}[1]{}
   \newcommand{\jianrong}[1]{}
   \newcommand{\jh}[1]{}
   \newcommand{\sooo}[1]{}
\newcommand{\sml[1]{}
\fi

\renewcommand{\eqref}[1]{Eq.\,(\ref{#1})}

\newcommand{\Sec}[1]{Sec.\,\ref{#1}}
\newcommand{\Secs}[2]{Secs.\,\ref{#1} and \ref{#2}}

\newcommand{\rcite}[1]{Ref.\,\cite{#1}}
\newcommand{\rrcite}[1]{Refs.\,\cite{#1}}

\newcommand{\NO}[1]{\,:\hspace{-1mm}{#1}\hspace{-1mm}:\,}

\def\mem{\hspace{0.1em}}
\def\hem{\hspace{0.05em}}
\def\nem{\hspace{-0.1em}}
\def\hnem{\hspace{-0.05em}}

\def\qiq{{\quad\implies\quad}}

\def\W{\mathcal{W}}

\def\b{\beta}

\def\t{\tau}
\def\bpsi{{\smash{\bar{\psi}}\kern0.02em\vphantom{\psi}}}

\def\db{{\dot{\b}}}

\def\lsq{{
    \kern-0.037em
    \adjustbox{scale=0.99,valign=c}{$
        {\lfloor \llap{\reflectbox{\rotatebox[origin=c]{180}{$\lfloor$}}}}
    $}
    \kern-0.04em
}}
\def\rsq{{
    \kern-0.04em
    \adjustbox{scale=0.99,valign=c}{$
        {\rlap{\reflectbox{\rotatebox[origin=c]{180}{$\rfloor$}}} \rfloor}
    $}
    \kern-0.037em
}}

\newcommand{\dbar}{
    d\kern-.20em\makebox[0pt][l]{$\bar{}$}\kern.20em
}
\newcommand{\deltabar}{
    \delta\kern-.20em\makebox[0pt][l]{$\bar{}$}\kern.20em
}

\newcommand{\BB}[1]{\Big(\,{#1}\,\Big)}
\newcommand{\bb}[1]{\bigg(\,{#1}\,\bigg)}
\newcommand{\bbsq}[1]{\bigg[\,{#1}\,\bigg]}
\newcommand{\bigbig}[1]{\big(\mem{#1}\mem\big)}
\newcommand{\lrp}[1]{\left(\,{#1}\,\right)}

\def\Kerr{{\smash{\text{$\kern-0.075em\sqrt{\text{Kerr\hem}}$}}}}

\def\lb{\{\kern-0.15em\{}
\def\rb{\}\kern-0.15em\}}

\newcommand{\act}[1]{[\,{#1}\,]}

\newcommand{\comm}[2]{[\hem{#1},{#2}\hem]}

\DeclareMathOperator{\ad}{ad}

\newcommand{\Texp}[1]{
    \mathrm{T}\kern-0.1em\exp\nem
    \bigg(\hem{
        #1
    }\bigg)
}

\renewcommand{\NO}[1]{{\mem{:}\mem{#1}\hem{:}\mem}}

\newcommand{\Ket}[1]{{\hem\big|\hem{#1}\big\rangle}}
\newcommand{\Bra}[1]{{\big\langle{#1}\hem\big|\hem}}

\newcommand{\expval}[1]{
	\big\langle\hem{
		#1
	}\hem\big\rangle
}

\let\oldexp\exp
\let\oldlog\log
\renewcommand{\exp}{\oldexp\nem}
\renewcommand{\log}{\oldlog\nem}

\usepackage{tikz}
\usetikzlibrary{matrix}
\usepackage{quiver}

\usetikzlibrary{calc} % to use relative coordinates
\usetikzlibrary{shapes.geometric} % to draw regular polygons
\usetikzlibrary{positioning} % to use right=of 
\usetikzlibrary{fit} % for fit size
\usepackage[a]{esvect} % arrow styling %f
\tikzset{empty/.style = {inner sep = 0pt, outer sep = 0, minimum size = 0}}
\tikzset{b/.style = {inner sep = 2pt, outer sep = 4pt, minimum size = 12pt}}
\tikzset{c/.style = {inner sep = 2pt, outer sep = 4pt, minimum size = 12pt}}
\tikzset{w/.style = {inner sep = 1pt, outer sep = 2pt, minimum size = 12pt, anchor = west}}
\tikzset{s/.style = {inner sep = 2.5pt, outer sep =2.5pt, minimum size = 1pt, font = \small}}
\tikzset{lin/.style = {draw, line width = 0.5pt}}

\def\arrow{\,\,\,\xrightarrow{\:\:\:}\,\,\,}

\usepackage{etoolbox}

\newcommand{\EikonalConvention}{New}

\newcommand{\EikonalSign}[1][]{%
  \ifdefstring{\EikonalConvention}{Old}{%
    \ifstrempty{#1}{-}{{\color{#1}-}}%
  }{%
  \ifdefstring{\EikonalConvention}{New}{%
  }{%
    \PackageError{EikonalSign}{Unknown option `\EikonalConvention'}%
      {Use Old or New.}%
  }}%
}
\newcommand{\MinusEikonalSign}[1][]{%
  \ifdefstring{\EikonalConvention}{New}{%
    \ifstrempty{#1}{-}{{\color{#1}-}}%
  }{%
  \ifdefstring{\EikonalConvention}{Old}{%
  }{%
    \PackageError{EikonalSign}{Unknown option `\EikonalConvention'}%
      {Use Old or New.}%
  }}%
}

\newcommand{\MinusEikonalSignC}{%
    \MinusEikonalSign[red]
}
\newcommand{\EikonalSignC}{%
    \EikonalSign[red]
}

\renewcommand{\EikonalSignC}{\EikonalSign}
\renewcommand{\MinusEikonalSignC}{\MinusEikonalSign}

\preprint{
\begin{flushright}
    \texttt{CALT-TH-2026-020}\\
    \texttt{CERN-TH-2026-115}\\
    \texttt{KIAS-P26031}\\
\end{flushright}
}

\title{The Diagrammar of Quantum Magnusian}

\author[a]{Li Guo}
\author[b]{Joon-Hwi Kim}
\author[c]{Jung-Wook Kim}
\author[d]{Sungsoo Kim}
\author[e]{Sangmin Lee}
\author[f]{Jian-Rong Li}
\affiliation[a]{Department of Mathematics and Computer Science,\\ 
Rutgers University, Newark, NJ 07102, U.S.A.}
\affiliation[b]{Walter Burke Institute for Theoretical Physics,\\
California Institute of Technology, Pasadena, CA 91125, U.S.A.}
\affiliation[c]{Theoretical Physics Department, CERN, 1211 Geneva 23, Switzerland}
\affiliation[d]{Department of Physics and Astronomy, Seoul National University, \\
1 Gwanak-ro, Gwanak-gu, Seoul 08826, Korea}
\affiliation[e]{School of Physics, Korea Institute for Advanced Study, \\
85 Hoegi-ro, Dongdaemun-gu, Seoul 02455, Korea}
\affiliation[f]{Faculty of Mathematics, University of Vienna,\\
Oskar-Morgenstern-Platz 1, Vienna, 1090, Austria}

\emailAdd{liguo@rutgers.edu,joonhwi@caltech.edu,jung-wook.kim@cern.ch,\\
sooo4017@snu.ac.kr,sangminlee@kias.re.kr,lijr07@gmail.com}

\abstract{
    The logarithm of the time-evolution operator has been termed Magnusian,
    on account of the fact that
    its expansion describes
    the Magnus series.
    The diagrammatic expansion and computation
    of the classical Magnusian have been completely established
    in terms of tree graphs and their Hopf algebra.
    Recent works initiated extensions
    into quantum field theory, 
    revealing general structures of loop expansions
    while finding intriguing relations between different diagrams.
    In this work, we advance the loop expansion further by providing 
    an efficient diagrammatic algorithm to calculate the weight factor of each graph in the quantum Magnusian,
    known as the Murua coefficient. 
    This is achieved by incorporating two complementary perspectives on the Magnusian
    at the same time:
    the color basis and the black-and-white basis.
    We extract the Murua coefficients from the Magnus series
    by utilizing these two bases
    while implementing
    an exponentiated Wick contraction.
    In turn, we identify
    the loop-level extension of Murua's recursive formula.
    Eventually, we establish a set of edge-contraction rules
    which
    facilitate a direct recursive computation
    of the Murua coefficients
    at the purely diagrammatic level,
    without referencing or directly manipulating 
    the underlying Magnus expansion.
    This shows that
    the matrix elements of the quantum Magnusian can be computed from graph manipulations alone. 
}

\begin{document}
\maketitle
\flushbottom

\bibliographystyle{utphys-modified}
\renewcommand*{\bibfont}{\footnotesize}
\setcounter{footnote}{0}

% \pagebreak
\section{Introduction}

The time-evolution operator of a closed quantum system
is represented by a unitary matrix. 
In the scattering setup,
the $S$-matrix is the unitary operator
encapsulating the time evolution
from far past to far future.
The unitarity of the $S$-matrix
can be made explicit by writing it as an exponential of a Hermitian matrix $\chi$: $S = e^{{\EikonalSignC}\chi/i\hbar}$.
This exponential representation
appeared in the early days of quantum field theory (QFT)~\cite{Lehmann:1957zz}
but was largely forgotten until recently~\cite{Damgaard:2021ipf,Damgaard:2023ttc}.
This recent revival was motivated by
modern applications of QFT techniques to gravitational-wave physics,
a field that has enjoyed exponential growth since the first direct detection of gravitational waves~\cite{LIGOScientific:2016aoc}.
The exponential representation has been shown to
simplify and clarify the classical limit of the $S$-matrix
and its use.

Most computations in QFT are performed in perturbation theory,
where 
the \textit{Dyson series} establishes the Feynman diagram expansion of the $S$-matrix.
The importance of diagrammatic tools for QFT computations cannot be overstated.
Historically, Feynman diagrams served as the very tool that
empowered Feynman to reproduce 
Slotnick's six-month-long calculation
within a single evening~\cite{feynman1965qed}. 
Besides the computational power,
diagrams serve as languages that are easily accessible to a large audience.
The current 
status of QFT---a universal tool for describing physics from the subnuclear scale of colliders to the gigaparsec scale of cosmic correlations---is %may be  
partly indebted to the accessibility of calculations provided by the diagrammatic expansions. 

While textbook expositions 
portray Feynman diagrams
as a derived construct from the Dyson series,
it is also possible to reverse the point of view, in fact.
In the celebrated paper titled
``Diagrammar''~\cite{DIAGRAMMAR},
't Hooft and Veltman raise the following question:
\emph{``Can QFT be formulated directly as an algebra of Feynman diagrams, with unitarity, causality, gauge invariance, renormalization, and field redefinitions treated as diagrammatic identities?''}
In this view,
perturbative QFT is practically an algebra of two-point functions
as per Wick's theorem.
It makes diagrams the starting point,
and then derives the needed formal properties from them.
It takes Feynman diagrams seriously as the primary language of perturbative QFT,
rather than merely as a mnemonic derived from the canonical formalism or the path integral formalism.

Considering this diagrammar discourse on the QFT $S$-matrix,
it is natural to question whether an analogous construction
exists for the logarithm of the $S$-matrix, $\chi = {\EikonalSignC}i\hbar\log S$.

At the \textit{tree level},
this question was answered in \rcite{KKKL};
see also \rrcite{Cristofoli:2021jas,Gonzo:2024zxo,Kim:2024grz} for related preceding works.
\rcite{KKKL} pointed out that
the logarithm of the $S$-matrix
is computed by the \textit{Magnus series}~\cite{Magnus},
which describes a characteristic nested bracket (commutator) structure
unlike the Dyson series.
By building upon this observation, 
it was shown that
the Magnus series
systematically derives
the diagrammatic rules for the classical limit of $\chi$ in terms of directed tree graphs.
A graph function $\omega (\tau)$,
known as the Murua coefficient~\cite{Murua}, 
assigns the correct weight when converting each directed tree graph $\t$ to its corresponding integrand.
The Murua coefficient $\omega(\tau)$ satisfies nontrivial relations referred to as \emph{edge-contraction rules} and \emph{sum rules}. 
Crucially,
these nontrivial relations can be used to ``bootstrap'' the values of $\omega(\tau)$ purely at the diagrammatic level,
precisely implementing the diagrammar philosophy.
To clarify,
this means that $\omega(\tau)$ can be computed
without 
making references to or
directly expanding out 
the Magnus series.

Through further developments in \rrcite{Kim:2025hpn,Alessio:2025flu,Kim:2025olv,Haddad:2025cmw,Magnusian,Kim:2025sey,Gonzo:2026yha},
%\jwb{ Haddad:2025cmw also needs to be cited but the flavour of this paper (in terms of terminology use) is somewhat different from the papers cited here}
it became clear that
this tree-level diagrammar
applies to both particles and fields,
incorporating conservative as well as dissipative dynamics.
It was also realized that
it universally applies to
not only $\chi = {\EikonalSignC}i\hbar \log S$
but also the logarithm of
any unitary time-evolution operator.
To emphasize this generality,
we adopt the term \emph{Magnusian}.
This term was first coined in \rcite{Magnusian}
for referring to the logarithm of unitary time-evolution operators in quantum mechanics. 

The diagrammatic exploration of
the \textit{loop-level} Magnusian
was initiated in \rrcite{Pichini,PSFOR-S},
offering complementary viewpoints on the same problem.

In \rcite{Pichini},
it was shown that the Murua coefficients $\omega(G)$
for generic graphs $G$
satisfy %a nontrivial relation:
the \emph{cutting rule}, which relates diagrams that differ only by cut propagators. 
%can be converted to the other by removing an edge. 
This relation alone, however, was not enough to uniquely determine the loop-level Murua coefficients at the diagrammatic level.
As a result, \rcite{Pichini}'s
determination of $\omega(G)$
still required directly manipulating and expanding out the Magnus series
(in terms of a useful reorganized formula),
unpacking nested commutators into operator products.
An explicit demonstration of this computation
was provided up to one-loop in $\phi^3$ theory,
together with a discussion on higher loop orders.

In \rcite{PSFOR-S},
it was shown that the frameworks of 
phase space formulation~\cite{Moyal:1949sk,Groenewold:1946kp,zachos2005quantum,Curtright:2011vw}
and deformation quantization~\cite{bayen1977quantum,kontsevich}
lead to a principled approach to the quantum Magnusian to all loop orders.
A concrete operational definition of the Murua coefficients $\omega(G)$ for all graphs $G$
was established
by utilizing the Moyal/Wick star product formalisms.
This formulation emphasizes and preserves 
the key feature
of the Magnus series,
i.e., the nested bracket structure.
The computation of $\omega(G)$
was demonstrated to all loops
at three vertices
with arbitrarily high degrees.
It was observed that
the tree-level coefficients
are recycled into a large portion of the loop-level coefficients,
leading to an idea dubbed fuzzification.

An approach that directly references and manipulates the Magnus series
could be called 
a ``first principles'' approach.
In contrast,
the diagrammar program pursues
a ``bootstrap'' approach
that seeks a purely graphical formulation and computation
of the Magnusian.
Although the former mode of computation may be efficient enough in practice,
it is still desirable to
establish the latter way of solving
and understanding the problem:
\begin{center}
    What is the \textit{diagrammar of the Magnusian}?
\end{center}

In this work,
we present a complete diagrammar for the quantum Magnusian.
To reiterate,
the major difference from the previous works~\cite{Pichini,PSFOR-S}
is that we focus on
the relations between Murua coefficients of graphs 
\textit{implied} by the Magnus series, instead of the Magnus series itself.

We utilize two distinct bases for 
diagrammatically representing the Magnusian:
color and black-and-white (BW).
These bases correspond to different choices of propagators (as two-point functions) for expressing the integrand. 

Firstly,
the color basis considered in \rcite{PSFOR-S} transparently encodes the algebraic structure of the Magnus series.
Consequently, the derivation of edge-contraction rules is relatively straightforward. 
Another advantage of this basis is that consistency of operator ordering imposes further constraints on the Murua coefficients, 
which provide enough information for computing the coefficients without directly expanding the Magnus series. 

Secondly,
the BW basis considered in \rcite{Pichini} manifests Lorentz invariance of the Magnusian, 
which is obscured in the color basis. 
The Hermiticity of the Magnusian also holds term by term in the BW basis, 
while it holds via complex conjugation in the color basis. 
Despite these differences,
both bases encode the same information and it is possible to switch between the bases.

The main technical novelty of this paper is the loop-level extension of the Murua formula~\cite{Murua}, 
which is a recursive algorithm to extract Murua coefficients from the Magnus recursion formula~\cite{Magnus} 
without actually evaluating the nested propagators. 
The original Murua formula was restricted to rooted tree graphs. 
It was then extended to all tree graphs in \rcite{KKKL}.  
This paper extends it to all loop graphs.
As an important corollary of the Murua formula, we derive edge-contraction rules 
in both the color basis and the BW basis. 
In practice, the contraction rules offer the fastest way to produce Murua coefficients.

The main body of this paper is divided into two sections. 
In \Sec{sec:Q-Mag}, we spell out the diagrammar of the quantum Magnusian.
The exposition itself may overlap with the two previous papers~\cite{Pichini,PSFOR-S}
to an extent.
The goal is
to establish our notations and connect the two complementary viewpoints of \rrcite{Pichini,PSFOR-S}, 
which we can utilize in computing the Murua coefficients. 
One novelty here is that we extend the notion of ``fuzzy propagator" of \rcite{PSFOR-S} 
to the BW basis of \rcite{Pichini}. The fuzzy propagator handles all ``banana loop" graphs 
and offers further insights into loop graphs.  
In \Sec{sec:Murua}, we extend the Murua formula to loop graphs in the two bases. 
The formula leads to the edge-contraction rules 
which turn into an efficient algorithm to compute the Murua coefficients for all loop graphs. 
A summary of the main results of \Sec{sec:Q-Mag} and \Sec{sec:Murua} can be found in \Sec{sec:summary-diagrammar}.
Finally, we conclude the paper in \Sec{sec:conclusion}
with a brief discussion on future directions. 

We provide a set of ancillary files in the companion repository \href{https://github.com/KIAS-Amplitudes/Quantum-Magnusian}{Quantum Magnusian}, where readers can look up the Murua coefficients and explore relations among graphs.

% \newpage 
\section{Quantum Magnusian}
\label{sec:Q-Mag}

We begin by quickly reminding ourselves of 
the basics of the Magnus expansion~\cite{Magnus}.

\subsection{Review of Magnus Series}
\label{sec:review}

The Magnus expansion concerns
a class of first-order differential equations
of the form
\begin{align}
    \label{1de}
    \dot{Y}(t) \,=\, h\, A(t)\, Y(t) 
    \,,\quad 
    Y(t_0) \,=\, Y_0
    \qiq
    Y(t) \,=\,
    e^{\Omega(t)}\hem
        Y_0
    \,,
\end{align}
where $A(t)$ is valued in a Lie algebra,
$Y(t)$ is vector-valued in a representation,
and $h$ is a formal parameter.
The Lie algebra can be either finite-dimensional or infinite-dimensional.
Since this describes a fairly generic setup,
the Magnus expansion finds broad applications in
diverse fields of mathematics and physics.

The original work by Magnus~\cite{Magnus} expresses
$\Omega(t)$ in \eqref{1de}
as the unique solution
to the following differential equation:
\begin{align}
    \dot{\Omega}(t) 
    \,=\,
        \frac{
            \operatorname{ad}_{\Omega(t)}
        }{
            e^{\operatorname{ad}_{\Omega(t)}} - 1
        }\act{
            A(t)
        }
    \,=\,
    A(t) \mem+\,
        \sum_{k \ge 1}\mem \frac{B_k}{k!}\mem
        \operatorname{ad}^k_{\Omega(t)}\act{
            A(t)
        }
    \,,\quad
    \Omega(t_0) \,=\, 0 
    \,.
\end{align}
Here, $\operatorname{ad}^k$ is the $k$-th iteration of
the adjoint action
$\operatorname{ad}_X\act{Y} := \comm{X}{Y}$,
while $B_n$ are the Bernoulli numbers
such that $B_1 = -1/2$.
Explicitly, the Magnus expansion is given by 
\begin{align}
    \Omega(t)
    \,=\,
    \sum_{n \geq 1}\,
        h^n\,
        \Omega_n(t)
    \,,\quad
    \Omega_1(t)
    \,=
        \int_{t_0}^t A(s)\, ds
    \,,
\end{align}
where $\Omega_{(n)}$ for $n \ge 2$
are obtained recursively by integrating
the following equation with the boundary condition
$\Omega_n(t_0) = 0$:
\begin{align}
\begin{split}
    \dot{\Omega}_n(t)
    \,=\,
        \sum_{k \ge 1}\,
            \frac{B_k}{k!}\,
        \sum_{\{r\}_k}\,
            \comm{
                \Omega_{r_k}\nem(t)
            }{
            \cdots
            \comm{
                \Omega_{r_2}\nem(t)
            }{
            \comm{
                \Omega_{r_1}\nem(t)
            }{
                A(t)
            }
            }
            \cdots
            }
    \,.
\end{split}
\label{Magnus-recursion}
\end{align}
Here, $\{r\}_k$ denotes the set of all $k$-tuples of integers satisfying 
\begin{align}
    r_1 + r_2 + \cdots r_k \,=\, (n-1)
    \,,\quad 
        r_i > 0
    \,.
\end{align} 
See, e.g., 
\rrcite{Blanes:2008xlr,Ebrahimi-Fard} for modern reviews on the Magnus expansion.
It is also known that
some formulas can express
$\Omega_n(t)$ directly
as a time-ordered integral of nested commutators of $A(t)$~\cite{Str,Pichini}.
We will not need any such formulas within this paper, 
although keeping its possibility in mind might help comprehend the diagrammatic expansion better. 

The Magnus expansion is readily applied to quantum mechanics and QFT 
with the commutator Lie algebra of quantum operators.
In a common physics notation,
the quantum Magnusian
in the interaction picture
is defined as
\begin{align} 
\label{def:quantum-Magnusian}
    {\EikonalSignC}
    \frac{1}{i\hbar}\,
        \chi
    \,=\,
        \log\bbsq{
            \mathcal{T} \exp\bb{
                \frac{1}{i\hbar} \int H_I(t) dt
            }
        }
    \,,
\end{align}
where $\mathcal{T}$ implements the time-ordered product,
and $H_I(t)$ is a Hamiltonian operator
in the interaction picture.\footnote{
    The sign convention in \eqref{def:quantum-Magnusian} is opposite to those of \rrcite{Damgaard:2021ipf,Damgaard:2023ttc,KKKL,Pichini}
    and follows \rrcite{Kim:2025sey,PSFOR-S}.
    This change of convention removes many unimportant minus signs in formulas. 
}

The formula in \eqref{Magnus-recursion} translates to
\begin{align}
\begin{split}
\label{Magnus-recursion-QM}
    {\EikonalSignC}
    \dot{\chi}_n(t)
    \,=\,
        \sum_{k \ge 1}\,
            \frac{B_k}{k!}
            \frac{1}{(i\hbar)^k}\,
        \sum_{\{r\}_k}\,
            \comm{
                \chi_{r_k}\nem(t)
            }{
            \cdots
            \comm{
                \chi_{r_2}\nem(t)
            }{
            \comm{
                \chi_{r_1}\nem(t)
            }{
                H_I(t)
            }
            }
            \cdots
            }
    \,.
\end{split}
\end{align}
As emphasized in \rcite{KKKL}, the $k$-fold nested commutator is accompanied by $(i\hbar)^k$ in the denominator, ensuring a well-defined classical limit. 

For practical use
of the quantum Magnusian,
one considers the in- and out-states as
\begin{align}
    \label{in-and-out}
    \Ket{\text{out}}
    \,=\,
        \mathcal{T} \exp\bb{
            \frac{1}{i\hbar} \int H_I(t) dt
        }
        \Ket{\text{in}}
    \,=\,
        e^{{\EikonalSignC}\chi/i\hbar}\mem
        \Ket{\text{in}}
    \,.
\end{align}
For a quantum observable $\mathcal{O}$,
it then follows that
\begin{align}
    \label{impulse}
    \expval{\mathcal{O}}_{\hnem\text{out}}
    \,=\,
        \Bra{\text{out}}
            \mathcal{O}
        \Ket{\text{out}}
    \,=\,
        \Bra{\text{in}}
            e^{{\MinusEikonalSignC}\chi/i\hbar}\mem
            \mathcal{O}\mem
            e^{{\EikonalSign}\chi/i\hbar}
        \Ket{\text{in}}
    \,=\,
        \expval{
            e^{{{\MinusEikonalSignC}\nem\ad_{\chi/i\hbar}}}\mem
            \mathcal{O}
        }_{\text{in}}
    \,.
\end{align}
Again, the classical limit of \eqref{impulse}
is well-defined~\cite{Kim:2024grz,KKKL}.
It leads to the so-called nested bracket formula~\cite{Kim:2024grz,Gonzo:2024zxo}
for computing the impulse of observables.

The main goal of this paper is to develop
a diagrammatic expansion of the quantum Magnusian, 
generalizing the tree result of \rcite{KKKL} 
as well as
the earlier works~\cite{Pichini,PSFOR-S}
on the loop case.

\subsection{Diagrammatic Expansion}

The diagrammatic expansion of the quantum Magnusian takes the schematic form, 
\begin{align}
\label{Mag-schematic}
    {\EikonalSignC}\chi
    \,\,\,\sim\,\,\,
        \sum_{G}\, 
            \frac{\omega(G)}{\sigma(G)}\,
            \hbar^{L(G)}\,
            \mathcal{I}(G) 
    \,. 
\end{align}
We shall explain all symbols entering the formula and fix the normalization factors, reviewing and 
combining similar definitions in \rrcite{Pichini,PSFOR-S}. 

The Magnusian is well-defined for any quantum system~\cite{PSFOR-S}, but in most parts of this paper, we will 
presume a scalar relativistic QFT as in \rcite{Pichini}. 

\paragraph{Graphs}

The sum in \eqref{Mag-schematic} runs over a set of graphs, which are similar but not identical to Feynman graphs.
Before describing the details, we define some basic quantities common to all graph expansions. 
For a graph $G$, we define 
\begin{align}
\begin{split}
    V(G) &\,=\, \mbox{(number of vertices of $G$)} 
    \,,\\
    E(G) &\,=\, \mbox{(number of edges of $G$)} 
    \,,\\
    L(G) &\,=\, \mbox{(number of loops of $G$)}
    \,.
\end{split}
\end{align}
They are subject to a relation 
\begin{align}
    E \,=\, L + V-1  \,.
    \label{graph-topology}
\end{align}
For comparison with \rcite{KKKL}, we will sometimes write $|G|$ to mean $V(G)$.  
Tree graphs ($L = 0$) will often be denoted by $\tau$ as in \rcite{KKKL}. 
As is customary in the QFT literature, we will use $\hbar$ as a formal loop counting parameter, 
as stated explicitly in \eqref{Mag-schematic}. 
The symmetry factor $\sigma(G)$ is integer-valued, on which we will elaborate shortly. 

The graph function $\omega(G)$ takes values in rational numbers. 
It was first introduced by Murua \cite{Murua} for rooted trees and
extended to non-rooted trees in \rcite{KKKL}. 
Some examples of $\omega(G)$ for loop graphs were given in \rrcite{Pichini,PSFOR-S}. 
We will generalize it to arbitrary loop graphs. How to compute $\omega(G)$ is the main technical novelty of this paper. 

\paragraph{Edges: Color vs. BW} 

\begin{figure}[htbp]
    \centering
    \includegraphics[width=0.4\linewidth]{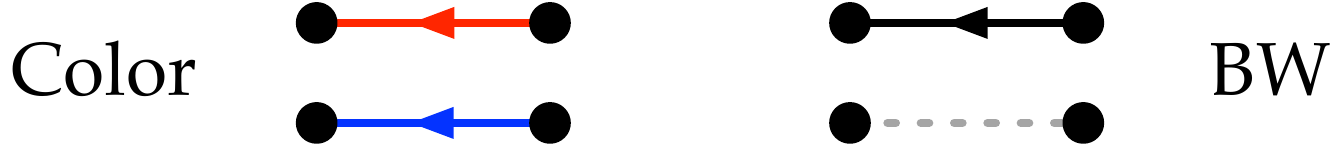}
    \caption{Edges in the color basis (left) and the BW basis (right).}
    \label{fig:basis-edges}
\end{figure}

The edges of our graphs denote propagators. 
We will use two distinct representations of edges:
the color basis of \rcite{PSFOR-S} 
and
the BW basis of \rcite{Pichini}. 
In the color basis, the edges are all directed and can be colored either red or blue. 
In the BW basis, the black edges are directed whereas the white (dotted gray for visualization) edges are undirected. 
The directions indicate time ordering. 
The black and white correspond to retarded and cut propagators, respectively.
We will specify the precise definitions of all propagators in \Sec{sec:propagators}. 
Fig.~\ref{fig:basis-edges} shows our graphical notation for the edges. 

\noindent
The number of edges splits into two types in each basis.  
\begin{align}
 \mbox{(Color):}\;\;   E \,=\, E_\text{red} + E_\text{blue} \,,
 \qquad 
  \mbox{(BW):}\;\;   E \,=\, E_\text{ret} + E_\text{cut} \,.
\end{align}

The schematic formula in \eqref{Mag-schematic} turns into specific formulas in the two bases:
\begin{subequations}
\label{Mag-both}
    \begin{align}
\label{Mag-color}
    {\EikonalSignC}\chi 
    \,&=\,
        \sum_{G_c}\,  
            \frac{\omega_c(G_c)}{\sigma_c(G_c)}\,
            \hbar^{L(G_c)}\,
        i^{V(G_c)-1}\, \mathcal{I}(G_c) 
    \\
\label{Mag-BW}
    \,&=\,
       \sum_{G_\text{bw}}\,
            \frac{\omega_\text{bw}(G_\text{bw})}{\sigma_\text{bw}(G_\text{bw})}\,
            \hbar^{L(G_\text{bw})}\,
        \mathcal{I}(G_\text{bw}) 
   \,. 
\end{align}
\end{subequations}
In the color basis, separating the overall factor of $i^{V-1}$---which originates from \eqref{Magnus-recursion-QM}---will turn out to be useful when 
discussing the Hermiticity of the sum. 
In the BW basis, the factor $i^{V-1}$ has been absorbed into the definition of the propagators such that Hermiticity holds term by term. 

\paragraph{Acyclic graphs} 

Either in the color basis or in the BW basis, 
the sum in \eqref{Mag-schematic} runs over all \emph{acyclic} graphs; 
the graphs should not contain any closed loop of directed edges. 
The directions mean time ordering, 
so a closed directed loop contains a factor like
\begin{align}
\label{time-loop}
    \Theta(t_1 {\mem-\,} t_2)\, \Theta(t_2 {\mem-\,} t_3)\, \Theta(t_3 {\mem-\,} t_1) 
    \,,
\end{align}
which vanishes upon integration over $t_i$ $(i=1,2,3)$.

A special case of a directed closed loop is the ``tadpole", 
an edge emanating from a vertex and terminating at the same vertex. 
The normal ordering of the operators in the vertices, to be specified below, excludes tadpole graphs. 

\paragraph{Symmetry factor} 

\begin{figure}[htbp]
    \centering
    \includegraphics[width=0.42\linewidth]{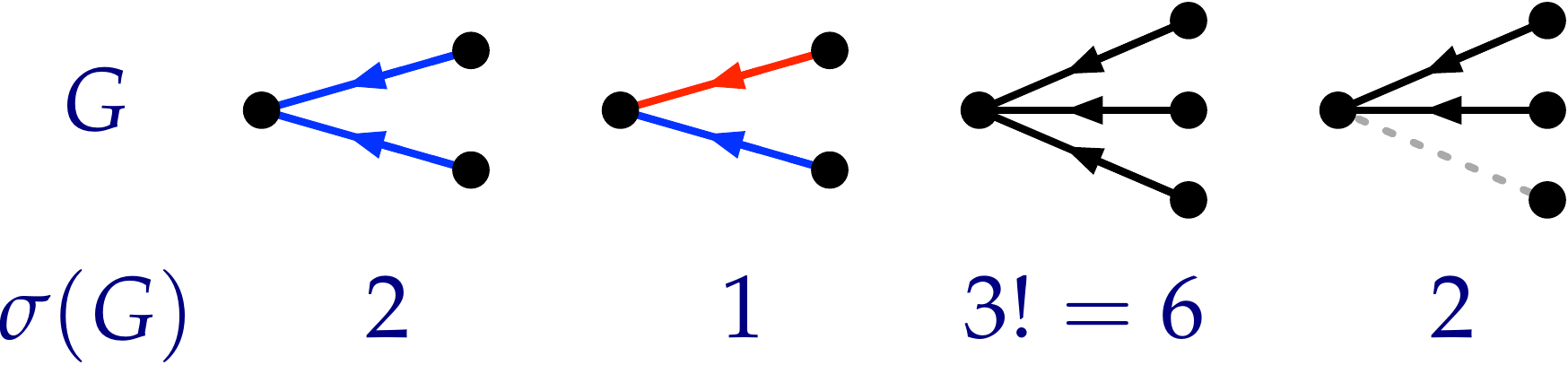}
    \caption{Symmetry factors of graphs.}
    \label{fig:symm-factor}
\end{figure}

The symmetry factor $\sigma(G)$ in \eqref{Mag-schematic} is defined in the standard way, 
which is completely independent of the specifics of the QFT. 
One minor concern is that 
the symmetry factor should take account of the directions and/or colors of the edges when applicable. 
See Fig.~\ref{fig:symm-factor} for some examples of the symmetry factor. 

\paragraph{Vertices} 

The vertices in the graphs denote copies of the interaction picture Hamiltonian $H_I$ in \eqref{def:quantum-Magnusian}. 
Quantum mechanically, they represent normal ordered operators. 
When we need a concrete theory, we will use the $\phi^4$ QFT, 
although the discussion is valid for any scalar potential, 
\begin{align}
    H_I(t) 
    \,=\,
    \int d^{d-1}\vec{x}\,\,
        \mathcal{V}(t,\vec{x}) 
    \,,\quad 
    \mathcal{V}(x)
    \,=\,
        \frac{\lambda}{4!} \NO{ \phi(x)^4 }
    \,.
\end{align}
Here, $\phi(x)$ is the free scalar field operator:
\begin{align}
\label{free-scalar-osc}
    \phi(x) 
    \,=
        \int \dbar^{d} k\, \deltabar(k^2{+\mem}m^2)\, \Theta(k^0)\mem
        \BB{
            a(k)\mem e^{ik\cdot x} 
            + a^\dagger(k)\mem e^{-ik\cdot x}
        }
    \,=\, 
        \phi^{+}(x) + \phi^{-}(x)
    \,.
\end{align}
Also, $\NO{\mathcal{O}}$ implements the normal ordering
on the operator $\mathcal{O}$,
which pushes all $a^\dagger$ to the left and all $a$ to the right.
When a vertex in a graph has valence $m$ ($m$ edges connected to the vertex), 
it contributes the $m$-th order derivative of $\mathcal{V}$. 
\begin{align}
    \mathcal{V} 
    \mem=\mem
        \frac{\lambda}{4!} \phi^4 
    \arrow
    \partial_\phi \mathcal{V} 
    \mem=\mem
        \frac{\lambda}{3!} \phi^3 
    \arrow
    \partial_\phi^2 \mathcal{V} 
    \mem=\mem
        \frac{\lambda}{2!} \phi^2
    \arrow
    \cdots 
    \,.
\end{align}
After all the vertices are connected by propagators, one takes the product of the remaining operators and 
takes an overall normal ordering. 
For example, the operator associated with a triangle graph is 
\begin{align}
\label{triangle-operator}
    \mathcal{O}_\triangle(x_1,x_2,x_3)
    \,=\,
    \lambda^3\mem
    \NO{
        \bb{ \frac{1}{2}\, \phi(x_1)^2 } 
        \bb{ \frac{1}{2}\, \phi(x_2)^2 }
        \bb{ \frac{1}{2}\, \phi(x_3)^2 }
    }
    \arrow
    \frac{\lambda^3}{2^3} 
    \NO{
        \phi_1^2 \phi_2^2 \phi_3^2
    }
    \,.
\end{align}
When we compute the matrix elements of $\chi$, the operator will act on the external bra/ket states. 

\paragraph{Operator-valued integrals}

Once we specify the vertices and the edges of a graph, we can write down the operator-valued integral $\mathcal{I}(G)$ in \eqref{Mag-color} or \eqref{Mag-BW}:
\begin{align}
\label{IG-OG-times-PG}
    \mathcal{I}(G) 
    \,=\,
    \int \prod_{i=1}^{|G|}\, d^dx_i\,\,
        \mathcal{O}_G(x_1, \cdots , x_{|G|})\,
        \mathcal{P}_G(x_1, \cdots , x_{|G|}) 
    \,.
\end{align}
The operator $\mathcal{O}_G$ generalizes the triangle example in \eqref{triangle-operator}. 
The other factor, $\mathcal{P}_G$, is the product of propagators 
whose precise definition will be given in \Sec{sec:propagators}.

%\newpage 
%%%%%%%%%%%%%%%%%%%%%%%%%%%%%%%%%%%%%%%%%%%%%%%%%%%%%%%
\subsection{Wick Contraction and Operator Product}
\label{sec:commutators}

A textbook approach to perturbative QFT uses the normal ordering prescription and the Wick contraction. 
It is useful to illustrate the key ideas with a toy model emulating 
the scalar field in \eqref{free-scalar-osc}, 
\begin{align}
\label{free-scalar-toy}
    \phi_i 
    \,=\, 
        \bar{f}_i\, a + f_i\, a^\dagger 
    \,,\quad 
    \comm{a}{a^\dagger} \,=\, \hbar
    \,,\quad 
    a \Ket{0} \,=\, 0  
    \,.
\end{align}
A simple exercise shows that 
\begin{align}
\label{toy-exercise} 
\begin{split}
    \phi_1 \phi_2 \; &= \; \NO{\phi_1 \phi_2} + \,\hbar\, \mathcal{W}_{12} \,,   
    \\
    \phi_1 \NO{\phi_2^2}  \; &= \; \NO{\phi_1 \phi_2^2} + 2\hbar\, \mathcal{W}_{12} \phi_2 \,,
    \\
    \phi_1 \phi_2 \phi_3 \; &= \; \NO{\phi_1 \phi_2 \phi_3} + \,\hbar\,  
    \bigbig{
        \mathcal{W}_{12} \phi_3 + \mathcal{W}_{23} \phi_1 + \mathcal{W}_{13} \phi_2
    }
    \,,
\end{split}
\end{align}
where $\mathcal{W}_{ij}$ is a toy Wightman function defined as
\begin{align}
    \hbar\mem \mathcal{W}_{ij} 
    \,=\,
        \hbar\, \bar{f}_i f_j 
    \,=\,
        \Bra{0} \phi_i \phi_j \Ket{0}
    \,. 
\end{align}
Let $\mathcal{P}(\phi)$ be a polynomial of $\phi$ and define $\mathcal{P}_i = \NO{\mathcal{V}(\phi_i)}$.
Then \eqref{toy-exercise} generalizes to  
\begin{align}
\label{Wick-bi-linear}
    \mathcal{P}_1 \mathcal{P}_2
    \,=\,
    \NO{ \mathcal{P}_1 \exp\left(
        \hbar% \mem \mathcal{W}_{12}
        \mem
        \frac{\overleftarrow{\partial}}{\partial a} 
        \frac{\overrightarrow{\partial}}{\partial a^\dagger} 
    \right)\mathcal{P}_2} = \NO{ \mathcal{P}_1 \exp
        \BB{\hbar\mem \mathcal{W}_{12} \overleftarrow{\partial_1} \overrightarrow{\partial_2}} \mathcal{P}_2}
    \,.
\end{align}
The first expression is universal and applicable to products with an arbitrary number of factors, confirming the associativity of the product, 
whereas the second expression is valid after specifying the two factors. 
It is understood that the partial derivatives here act as if $\phi_i$ were classical variables. 
Only after taking derivatives does one reinstate the operator nature of $\phi_i$. 
For later convenience, we give a name ``Wick operator" to the formal object 
\begin{align}
\label{Wick-op}
    W_{ij} \,=\, 
        \mathcal{W}_{ij}\mem
        \overleftarrow{\partial_i} 
        \overrightarrow{\partial_j}
    \,.
\end{align}

As recently elucidated by \rcite{PSFOR-S},
the cleanest and most straightforward way to
derive and understand
the exponential formula in \eqref{Wick-bi-linear}
is through the theory of star products~\cite{Moyal:1949sk,Groenewold:1946kp}.
See also \rcite{deformation-quantification}
for a path integral origin.
In this paper, however,
we choose to not invoke the star product formalism explicitly 
and
will continue with the standard language of textbook operator formalism
as in \eqref{free-scalar-toy}.

The exponentiation in \eqref{Wick-bi-linear} is of utmost importance throughout this paper. 
If we focus on one of the two derivatives, 
the Leibniz rule reincarnates as the identity
\begin{align}
    \exp\mem(c\, \partial_x)\act{
        f(x)\mem g(x)
    }
    \,=\,
        f(x {\,+\,} c)\mem g(x {\,+\,} c)
    \,=\,
        \BB{
            \exp\mem(c\, \partial_x) f(x)
        }\BB{
            \exp\mem(c\, \partial_x) g(x)
        } 
    \,.
\end{align} 
See \rcite{PSFOR-S} for an algebraic account on this identity.
In the diagrammatic expansion, the exponential form 
will allow us
to keep track of multiple propagators emanating from the same vertex with little effort.

\paragraph{Operator products}

The Magnus expansion involves nested commutators, 
\begin{align}
\begin{split}
\label{star-brackets-levels}
    \mbox{Level 1:} &\quad 
        \comm{\Omega_0}{\Omega_1}
    \,,\\
    \mbox{Level 2:} &\quad 
        \comm{\comm{\Omega_0}{\Omega_1}}{\Omega_2}
    \,,\\
    \mbox{Level 3:} &\quad 
        \comm{\comm{\comm{\Omega_0}{\Omega_1}}{\Omega_2}}{\Omega_3}
    \,. 
\end{split}
\end{align}
Here, $\Omega$ can be any composite operator; it can be the vertex factor $\mathcal{V}$ or 
more complicated objects as in \eqref{Magnus-recursion-QM}. 
The level-$k$ bracket produces $2^k$ products of operators, but the operator ordering of vertices and the sign of each term are well known from the outset, 
\begin{align}
\begin{split}
    \label{commutator-expanded}
    \comm{\Omega_0}{\Omega_1}
    \,&=\,
        \Omega_0 \Omega_1 - \Omega_1 \Omega_0
    \,,\\
    \comm{\comm{\Omega_0}{\Omega_1}}{\Omega_2}
    \,&=\,
        \Omega_0 \Omega_1  \Omega_2 + \Omega_2  \Omega_1 \Omega_0 - \Omega_1 \Omega_0 \Omega_2 - \Omega_2 \Omega_0 \Omega_1 
    \,. 
\end{split}
\end{align}
Therefore, we can recover the whole expression from ``monotonically increasing" products
of the form
\begin{align}
\label{product-Vn}
  \Omega_0  \Omega_1 \cdots \Omega_k 
  \,.
\end{align}
At $k=1$, we introduce a graphical representation, 
\begin{align}
    \label{notation-ds}
   \Omega_0  \Omega_1 
   \,\,=\,\,
   \,
   \raisebox{-2.4mm}{\includegraphics[width=7.6cm]{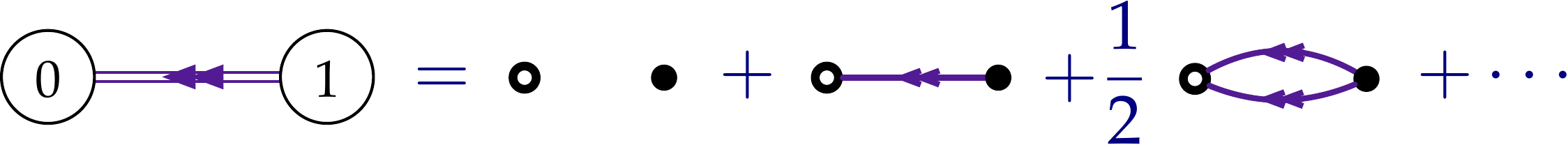}}
   \,
   \,,
\end{align}
which represents the operator product.
Namely, we draw double-stroke arrows to express the exponentiated Wick operator in \eqref{Wick-bi-linear};
this is an imitation of
\rcite{PSFOR-S}'s fuzzy line notation.

\eqref{notation-ds} illustrates that
a double-stroke line unpacks into
a series of graphs
by expanding the exponential.
In particular,
the leading part of this series
describes disconnected vertices.
It should be clear that
this part gets
canceled when one forms the commutator combinations
on the right-hand sides in \eqref{commutator-expanded}.

The two vertices can be connected by one or more Wightman functions, 
each of which is denoted by a single-stroke arrow. 
The ordering of $\mathcal{W}_{01}$ is specified by the arrow going from label 1 to label 0. 
Inspired by \rcite{PSFOR-S}, to distinguish the operator ordering from the time ordering, 
we are using doubled arrowheads for the former. 

\begin{figure}[htbp]
    \centering
    \includegraphics[width=0.84\linewidth]{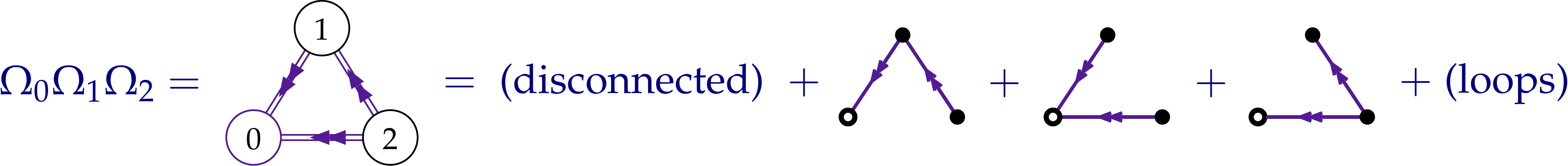} 
    \\
     \includegraphics[width=0.7\linewidth]{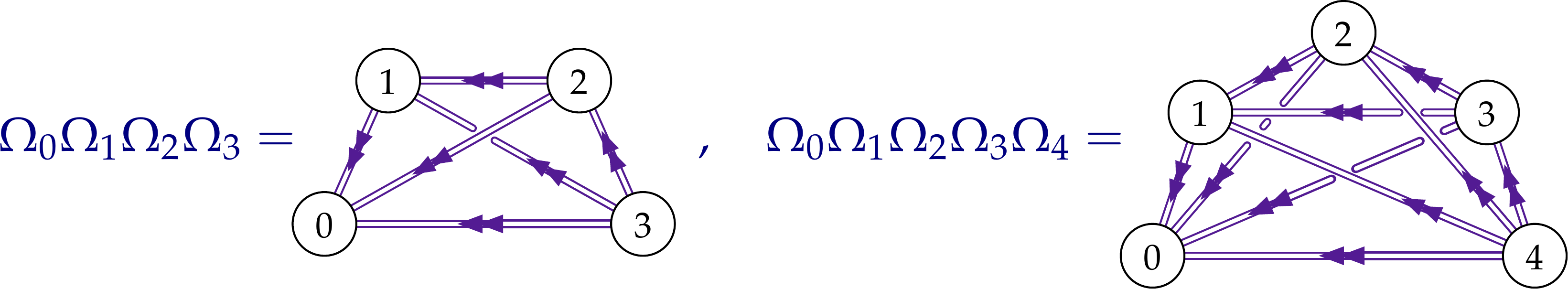}
    \caption{Operator product at level 2.}
    \label{fig:spider-2}
\end{figure}

Fig.~\ref{fig:spider-2} shows the operator product at $k=2, 3, 4$. 
The arrows always go from a bigger integer label to a smaller one. 
The associativity of the product,  $(\Omega_0 \Omega_1 )\Omega_2 = \Omega_0 (\Omega_1 \Omega_2)$, 
is recognizable in the figure. 

It is clear how the pattern generalizes to the level $k$ product in \eqref{product-Vn}. 
At level $k$, there are $(k+1)$ vertices. 
Vertex with label $j$ ($0\le j \le k$) has $j$ outgoing double-stroke arrows. 
The double-stroke (exponentiated) arrows are expanded in single-stroke arrows.

Disconnected diagrams all cancel out when they participate in the commutators in \eqref{commutator-expanded}. 
Connected diagrams from the operator products, 
dressed with the signs from the nested commutators 
and the time-ordering $\Theta$ functions, will 
form the building blocks of the graph sum for the Magnusian.

%\newpage 
%%%%%%%%%%%%%%%%%%%%%%%%%%%%%%%%%%%%%%
\subsection{Propagators} 
\label{sec:propagators}

We carry over the lessons learned from the toy model to a relativistic scalar QFT. 

\paragraph{Wightman and related functions} 

\begin{figure}[htbp]
    \centering
    \includegraphics[width=0.36\linewidth]{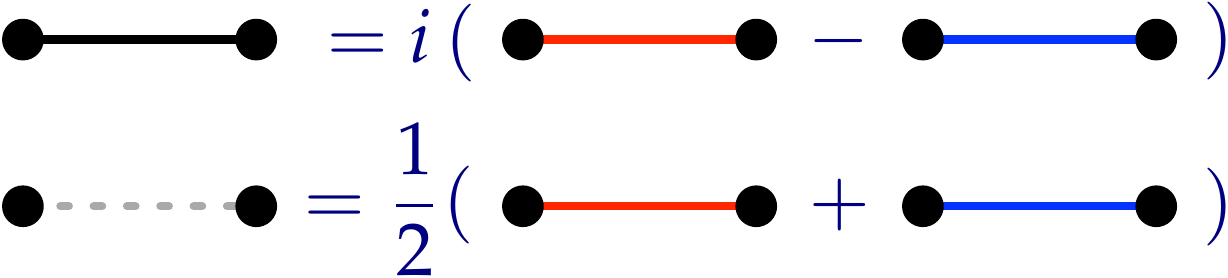}
    \caption{Basis change for Wightman functions}
    \label{fig:trans-basis}
\end{figure}

Our convention for the Wightman function is 
\begin{align}
    \W(x_1,x_2) \,= 
        \int \dbar^d k\,\,
            \deltabar(k^2{+\mem}m^2)\,
            \Theta(k^0)\,
            e^{ik\cdot (x_1-x_2)} 
    \,=\,
        \Bra{0}
            \phi^{+}(x_1) \phi^{-}(x_2) 
        \Ket{0}
    \,. 
\end{align}
This is a homogeneous solution to the Klein-Gordon equation. 
As in \rcite{PSFOR-S}, 
it is useful to distinguish between the two orderings of a pair of spacetime points by colors, 
\begin{subequations}
   \begin{align}
    (\W_\text{\textcolor{red}{red}})_{12} 
    \,&=\,
        \W(x_1,x_2)
    \,,\\
    (\W_\text{\textcolor{blue}{blue}})_{12}
    \,&=\,
        \W(x_2,x_1) 
    \,=\,
        [\W(x_1,x_2)]^* 
    \,. 
\end{align} 
\end{subequations}
The symmetric (Hadamard) and anti-symmetric (Pauli-Jordan) linear combinations are equally useful, 
\begin{subequations}
\begin{align}
    (\W_S)_{12}
    \,&=\,
        \frac{1}{2}\,
        \BB{
            (\W_\text{\textcolor{red}{red}})_{12} + (\W_\text{\textcolor{blue}{blue}})_{12}
        }
    \,,\\
    (\W_A)_{12} 
    \,&=\,
        i\mem\BB{
            (\W_\text{\textcolor{red}{red}})_{12} - (\W_\text{\textcolor{blue}{blue}})_{12} 
        }
    \,. 
\end{align} 
\end{subequations}
See Fig.~\ref{fig:trans-basis} for a pictorial representation of the basis change for Wightman-like functions.

It is common to use the terms ``propagator" and ``Green's function" synonymously. 
We abuse the terminology somewhat: our ``propagators" below, except the retarded one, are 
not Green's functions. 

\begin{figure}[htbp]
    \centering
    \includegraphics[width=0.72\linewidth]{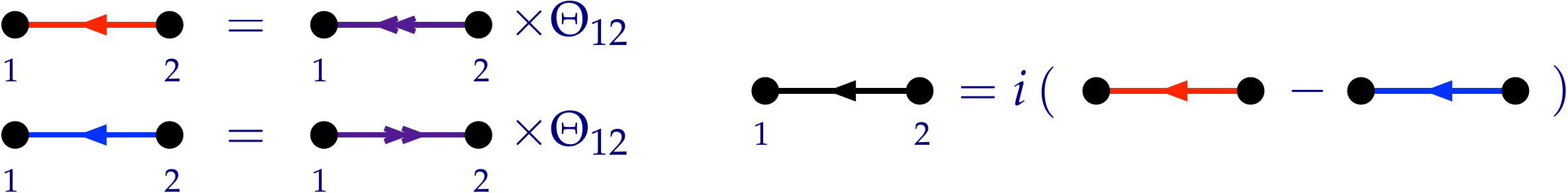}
    \caption{Propagators, with the distinction between time ordering and operator ordering.}
    \label{fig:prop-all}
\end{figure}

We define the colored propagators as\footnote{These functions were referred to as \emph{retarded Wightman functions} in App.~C of \rcite{KKKL}.} 
\begin{subequations}
\label{Color-prop}
    \begin{align}
    \label{red-prop}
    (G_\text{\textcolor{red}{red}})_{12} \,&=\,  
    (\W_\text{\textcolor{red}{red}})_{12} \,\Theta(t_1{-\,}t_2) \,,
    \\
    \label{blue-prop}
     (G_\text{\textcolor{blue}{blue}})_{12} \,&=\, 
     (\W_\text{\textcolor{blue}{blue}})_{12} \,\Theta(t_1{-\,}t_2) \,. 
\end{align}
\end{subequations}
The more familiar retarded and cut propagators are 
\begin{subequations}
\label{BW-prop}
    \begin{align}
    \label{ret-prop}
    (G_\text{ret})_{12} \,&=\,  (\W_A)_{12} \,\Theta(t_1{-\,}t_2) = i \left[ (G_\text{\textcolor{red}{red}})_{12} - (G_\text{\textcolor{blue}{blue}})_{12}\right] \,,
    \\
    \label{cut-prop}
     (G_\text{cut})_{12} \,&=\,  (\W_S)_{12} \,. 
\end{align}
\end{subequations}
Fig.~\ref{fig:prop-all} shows the graphical notation for the propagators, 
where the time ordering is denoted by single arrowheads while the operator ordering is denoted by double arrowheads. 
While the Wightman-like functions (\emph{undirected} red/blue/black edges) are useful when we discuss the basis change, 
they will enter our formulas for the Magnusian only through the propagators, \eqref{Color-prop} or \eqref{BW-prop}. 

Compared to \rcite{Pichini}, the cut propagator is scaled by a factor of 2. 
Our sign convention for the retarded propagator is the same as in \rcite{KKKL} 
and opposite to that of \rcite{Pichini}. 
As a consequence, the $\omega$ coefficients are related as $\omega_\text{here} = (-1)^{E_\text{ret}} \omega_\text{there}$. 
The factors of $(\pm i)$ are inserted such that under complex conjugation the propagators satisfy 
\begin{align}
\label{prop-cc}
[(G_\text{\textcolor{red}{red}})_{12}]^* = (G_\text{\textcolor{blue}{blue}})_{12} \,,
\quad 
    [(G_\text{ret})_{12}]^* = (G_\text{ret})_{12} \,,
    \quad 
    [(G_\text{cut})_{12}]^* = (G_\text{cut})_{12} \,.
\end{align}
%

%%%
\paragraph{Primary vs. descendant graphs} 

\begin{figure}[htbp]
    \centering
    \includegraphics[width=0.82\linewidth]{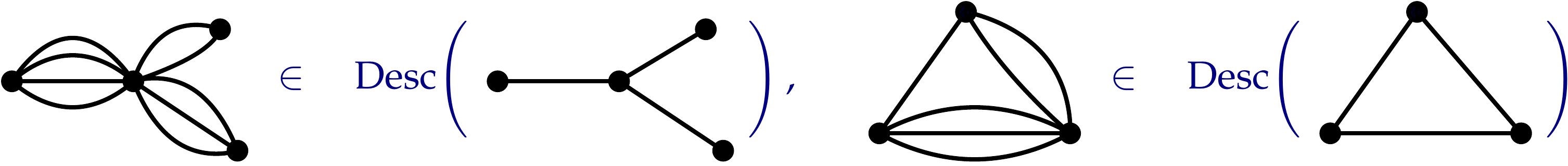}
    \caption{Primary graphs and their descendants.}
    \label{fig:primary-desc}
\end{figure}

When discussing graphs with loops, it is useful to distinguish ``genuine loops" involving three or more vertices 
from ``banana loops" consisting of multiple edges between two vertices. 
In what follows, graphs containing no banana loop will be called ``primary", and those containing some banana loops ``descendant"~\cite{PSFOR-S}.
See Fig.~\ref{fig:primary-desc} for an illustration. 

\paragraph{Fuzzy colored propagators} 

As explained in \rcite{PSFOR-S}, 
the exponentiated Wick operator in \eqref{Wick-bi-linear} and \eqref{Wick-op} incorporates the banana loops systematically. 
To see how it affects the Magnus expansion, consider the formal power series, 
\begin{align}
\label{color-fuzzy} 
\begin{split}
     \hbar W^{\#}_\text{\textcolor{red}{red}} \,&=\, e^{\hbar W_\text{\textcolor{red}{red}}} 
    \,=\, 1 + \hbar W_\text{\textcolor{red}{red}} + \frac{1}{2}\, (\hbar W_\text{\textcolor{red}{red}})^2 + \cdots\,, 
    \\
    \hbar W^{\#}_\text{\textcolor{blue}{blue}} \,&=\, e^{\hbar W_\text{\textcolor{blue}{blue}}} 
    \,=\, 1 + \hbar W_\text{\textcolor{blue}{blue}} + \frac{1}{2}\, (\hbar W_\text{\textcolor{blue}{blue}})^2 + \cdots\,. 
\end{split}
\end{align}
The coefficient $1/(k!)$ for the $k$-th order term accounts for the symmetry factor of the banana loop. 
Aside from the symmetry factor, the descendant graphs containing banana loops share the same $\omega(G)$ as the primary graph containing no banana loop. 

\begin{figure}[htbp]
    \centering
    \includegraphics[width=0.45\linewidth]{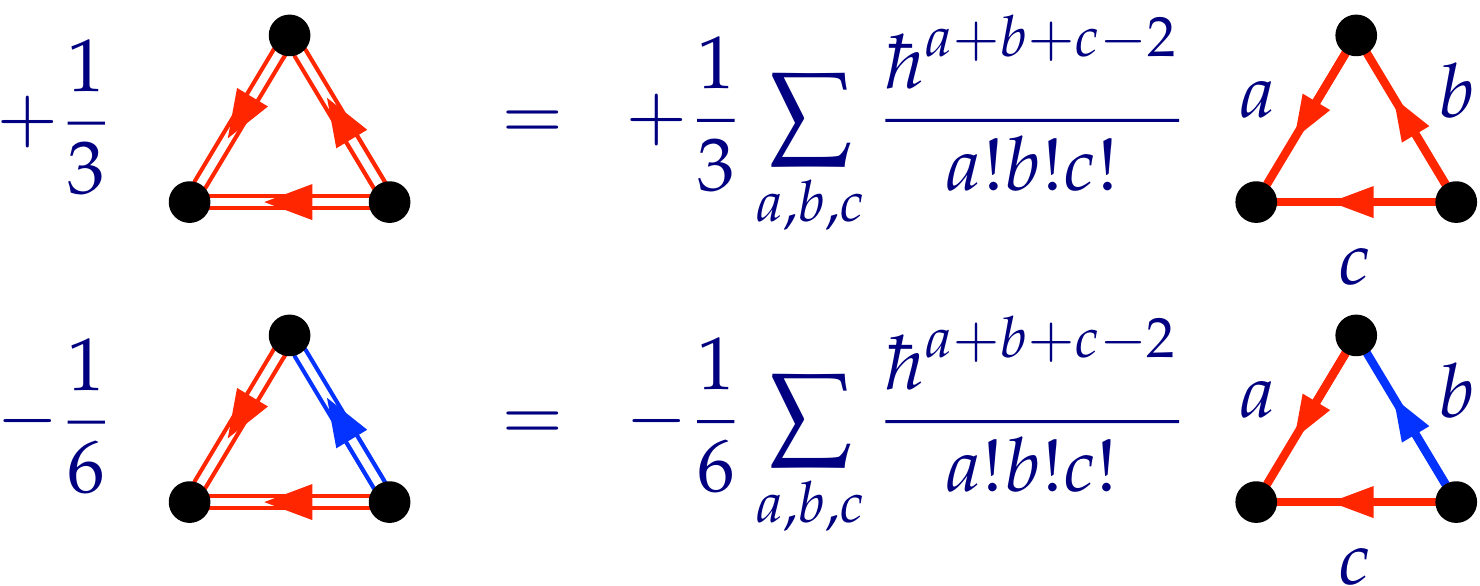}
    \caption{Fuzzy propagators account for all descendant graphs.}
    \label{fig:fuzzy-color}
\end{figure}

We treat \eqref{color-fuzzy} as a formal object, since $W$ in the expressions carries not only the Wightman functions $\mathcal{W}$ but also the functional derivatives acting on the field variable $\phi$. 
However, the combinatorics of the coefficients follow the simple rules mentioned above.  
Fig.~\ref{fig:fuzzy-color} reproduces simple examples of fuzzification from \rcite{PSFOR-S}.

%\newpage 
\paragraph{Fuzzy BW propagators} 

\begin{figure}[htbp]
    \centering
    \includegraphics[width=0.75\linewidth]{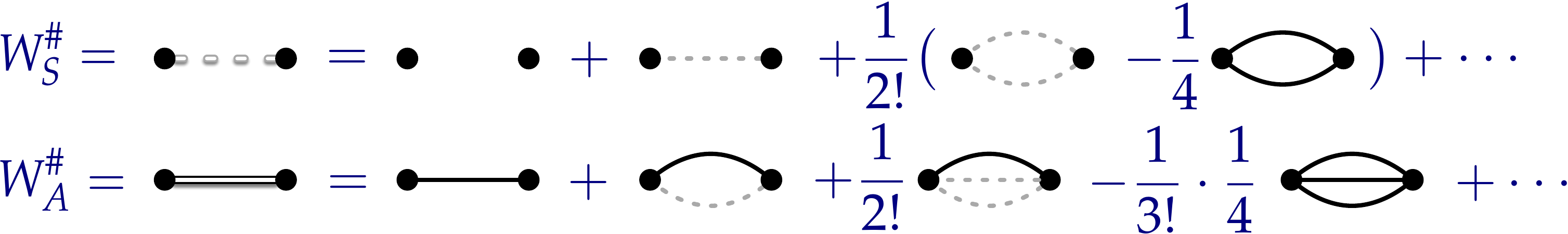}
    \caption{Fuzzy Wick operators in the BW basis.}
    \label{fig:fuzzy-bw}
\end{figure}

As before, the transition from the color basis to the BW basis 
is taking the symmetric or anti-symmetric linear combinations. 
The symmetric one is 
\begin{subequations}
\begin{align}
\label{bw-fuzzy-S}
    \hbar W_S^{\#} 
    \,&:=\,
    \frac{1}{2}\,\BB{
        e^{\hbar W_\text{\textcolor{red}{red}}} + e^{\hbar W_\text{\textcolor{blue}{blue}}}
    }
    \\
    \nonumber
    \,&=\,
        \frac{1}{2}\,\bb{
            \exp\bb{
                \hbar W_S -i\mem \frac{\hbar W_A}{2}
            }
            +
            \exp\bb{
                \hbar W_S +i\mem \frac{\hbar W_A}{2}
            }
        }
    \\
    \nonumber
    \,&=\,
        \sum_{m=0}^\infty\sum_{r=0}^{\lfloor m/2 \rfloor}
            \frac{\hbar^m (-1)^r}{(2r)!(m{\,-\,}2r)!}\,
            W_S^{m-2r}
            \bb{
                \frac{W_A}{2}
            }^{\nem\nem2r}
    \\
    \nonumber
    \,&=\,
        1+ \hbar W_S 
        + \frac{1}{2!}\, \hbar^2\mem \bb{
            W_S^2 - \frac{1}{4}\, W_A^2
        }
        + \hbar^3\mem \bb{
            \frac{1}{3!}\, W_S^3 
            - \frac{1}{2!}\, W_S \cdot \frac{1}{4}\, W_A^2
        } 
        + \mathcal{O}(\hbar^4)
    \,.
    \nonumber
\end{align}
The anti-symmetric one is 
\begin{align}
\label{bw-fuzzy-A}
    \hbar W_A^{\#} 
    \,&:=\,
    i\,\BB{
        e^{\hbar W_\text{\textcolor{red}{red}}} - e^{\hbar W_\text{\textcolor{blue}{blue}}}
    }
    \\
    \,&=\,
        i\,\bb{
            \exp\bb{
                \hbar W_S -i\mem \frac{\hbar W_A}{2}
            }
            -
            \exp\bb{
                \hbar W_S +i\mem \frac{\hbar W_A}{2}
            }
        }
    \nonumber
    \\
    \,&=\,
        W_A
        \sum_{m=1}^\infty\sum_{r=0}^{\lfloor (m-1)/2 \rfloor}
            \frac{\hbar^m (-1)^r}{(2r{\,+\,}1)!(m{\,-\,}2r{\,-\,}1)!}\,
            W_S^{m-2r-1}
            \bb{
                \frac{W_A}{2}
            }^{\nem\nem2r}
    \nonumber
    \\
    \,&=\,
        1+ \hbar W_A
        + \hbar^2\mem \BB{
            W_S W_A
        }
        + \hbar^3\mem \bb{
            -\frac{1}{3!} \cdot \frac{1}{4}\, W_A^3 
            +\frac{1}{2!}\, W_S^2 W_A
        } 
        + \mathcal{O}(\hbar^4)
    \,.
    \nonumber
\end{align}
\end{subequations}
In each term in the two expansions, we specified separately the symmetry factors and the powers of 2 from $\mathcal{W}_A/2$.
See Fig.~\ref{fig:fuzzy-bw} for a pictorial representation of the expansions. 

\begin{figure}[htbp]
    \centering
    \includegraphics[width=0.75\linewidth]{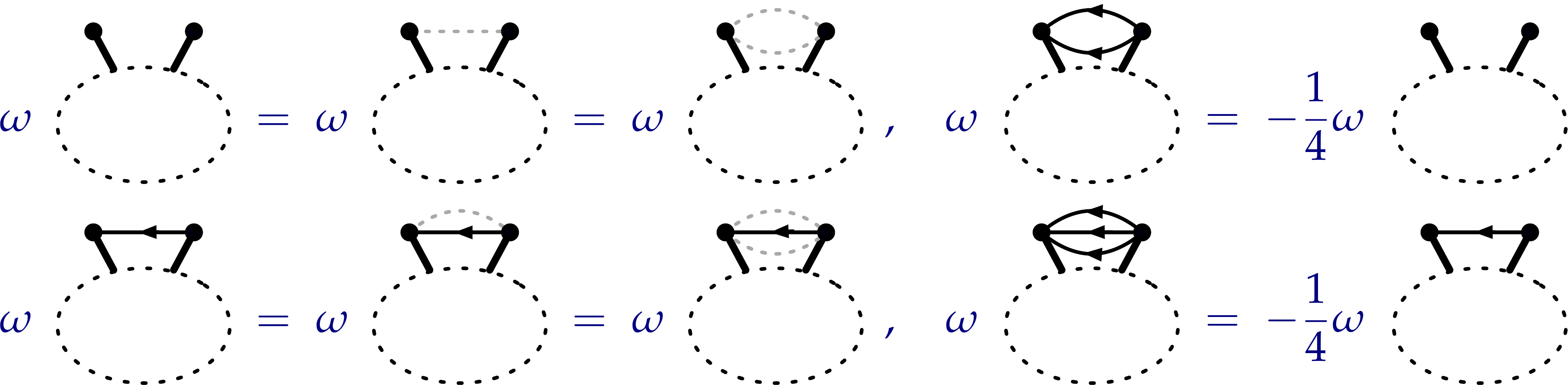}
    \caption{Implication of fuzzy propagators on the $\omega$ values}
    \label{fig:bw-omega-exp}
\end{figure}

We claim that the relative factors in the expansion imply relative normalization of the $\omega$ values of graphs descending from the same primary graph. 
In the descendant graphs, factors of $W_S$ only affect the symmetry factor and not the $\omega$ value. 
For $W_A$, in addition to the symmetry factor, we should multiply the $\omega$ value by $(-1/4)$ when a pair of $W_A$ is removed. 
See Fig.~\ref{fig:bw-omega-exp} for an illustration. 

%\newpage 
\subsection{Symmetry}

\paragraph{Time ordering and Lorentz invariance}

All $\W$-type functions are Lorentz invariant, but when they are multiplied by the time-ordering function $\Theta(t_i - t_j)$, Lorentz invariance may get broken. 
For example, the red and blue propagators in \eqref{Color-prop} are \emph{not} individually Lorentz invariant, since $\W$ does not necessarily vanish when $(x_1-x_2)$ is space-like. 
In contrast, both propagators in \eqref{BW-prop} are Lorentz invariant. 
For the retarded propagator, 
\begin{align}
    (G_\text{ret})_{12}
    \,=\,
        i\, \BB{
            (\W_\text{\textcolor{red}{red}})_{12} - (\W_\text{\textcolor{blue}{blue}})_{12}
        } \,\Theta(t_1{\mem-\,}t_2) 
    \,, 
    \label{ret-prop-recap}
\end{align}
the anti-symmetric combination cancels out the value for space-like $(x_1-x_2)$ completely. 
The $\Theta$ function selects the future light-cone and discards the past light-cone, 
which respects Lorentz invariance. 
The cut propagator does not carry a $\Theta$ function. 

When the colored propagators are combined to reach a cut propagator, the $\Theta$ functions should all be removed. 
There are broadly two mechanisms. One is 
\begin{align}
    (\W_S)_{12}\,
    \bigbig{
        \Theta_{12} + \Theta_{21}
    }
    \,=\,
        (\W_S)_{12} 
    \,. 
    \label{cut-prop-recap}
\end{align}
The other is 
\begin{align}
    (\W_S)_{12} \Theta_{12}
    \,
    (\W_A)_{13} \Theta_{13}
    \,
    (\W_A)_{32} \Theta_{32} 
    \,=\,
        (G_\text{cut})_{12}\mem
        (G_\text{ret})_{13}\mem
        (G_\text{ret})_{32}
    \,.
\end{align}
Here, the product $\Theta_{13} \Theta_{32}$ enforces $t_1 > t_2$, so $\Theta_{12}$ can be discarded. 

\paragraph{Time reversal and color parity} 

The diagrammatic expansion is invariant under two symmetry transformations. % at least 
The first one is time reversal. The expansion is invariant under the overall reversal of time ordering. 
Time-reversal images share the same $\omega$ values, so the sum over all graphs is invariant. 

The second symmetry is the ``color parity"; flipping all reds to blues and vice versa. 
It originates from the fact that $\chi$ is Hermitian (self-adjoint). 
For a graph $G$ in the color basis, let $\mathcal{C}(G)$ be the color parity image. 
The color parity acts on the $\omega$ value as 
\begin{align}
    \omega(\mathcal{C}(G)) 
    \,=\,
        (-1)^{V(G)-1} \omega(G)
    \,.
\end{align}
With the $i^{(V(G)-1)}$ factor in \eqref{Mag-color}, and the complex conjugation in \eqref{prop-cc}, this rule ensures that $\chi$ is Hermitian. 
The color parity translated to the BW basis implies that 
\begin{align}
\label{color-parity-BW}
    L \,-\mem E_\text{cut} 
    \,=\,  
        E_\text{ret} - (V {\mem-\,} 1) 
    \;\; \in \;\; 
        2\mathbb{Z} 
    \,. 
\end{align}
This condition (called ``cut structure" in \rcite{Pichini}) is necessary for the $i$ factors to be absorbed into the retarded propagators in going from the color basis to the BW basis. 

\subsection{Examples}

Before we explain systematic methods to compute $\omega(G)$ for an arbitrary graph, to familiarize the reader with the outcome, 
we present a few more examples both in the color basis and in the BW basis, 
and state some properties of $\omega(G)$ to be proved in later sections.

Throughout this paper, we will switch between two complementary viewpoints. One is to compute some results in the color basis 
and carry them over to the BW basis. 
The other is to figure out how to compute the $\omega$ values directly in the BW basis. 

\paragraph{2-point and 3-point}

\begin{figure}[htbp]
    \centering
    \includegraphics[width=0.64\linewidth]{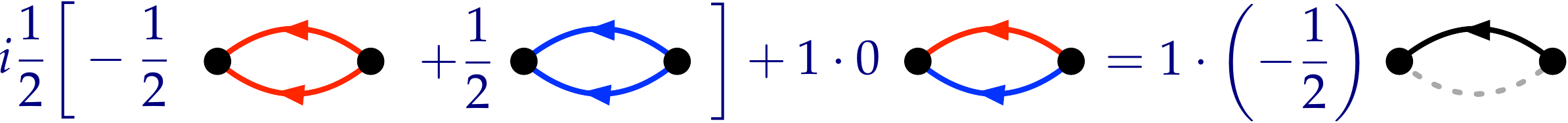}
    \caption{Color basis to BW basis: bi-angle.}
    \label{fig:trans-biangle}
\end{figure}

The simplest example of the color-to-BW map is shown in Fig.~\ref{fig:trans-biangle}. 
It encodes graphically the following identity.
\begin{align}
    i\, \frac{1}{2}\mem \bigbig{
        \mathcal{W}_\text{\textcolor{red}{red}}^2 - \mathcal{W}_\text{\textcolor{blue}{blue}}^2
    }_{\nem12} \Theta_{12} 
    \,=\,
        (G_\text{ret})_{12}\mem
        (G_\text{cut})_{12}
    \,.
\end{align}
In the color basis, the fuzzy propagator assigns the same $\omega$ value to the tree graph and the one-loop bi-angle graph. 
The parallel statement in the BW basis is that the graph on the RHS of Fig.~\ref{fig:trans-biangle} shares the same $\omega = -1/2$ as the 2-point tree without the cut propagator. 
This is the simplest case of the observation made in \rcite{Pichini}; the one-loop graphs with a single cut propagator share the same $\omega$ value as the corresponding tree graphs obtained by erasing the edge.  

We also observe how the color parity operates in the BW basis. 
According to \eqref{color-parity-BW}, 
the number of retarded propagators should be odd. 
For a loop with only two propagators, 
the only possibility is to have one retarded propagator and one cut propagator.

\begin{figure}[htbp]
    \centering
    \includegraphics[width=0.56\linewidth]{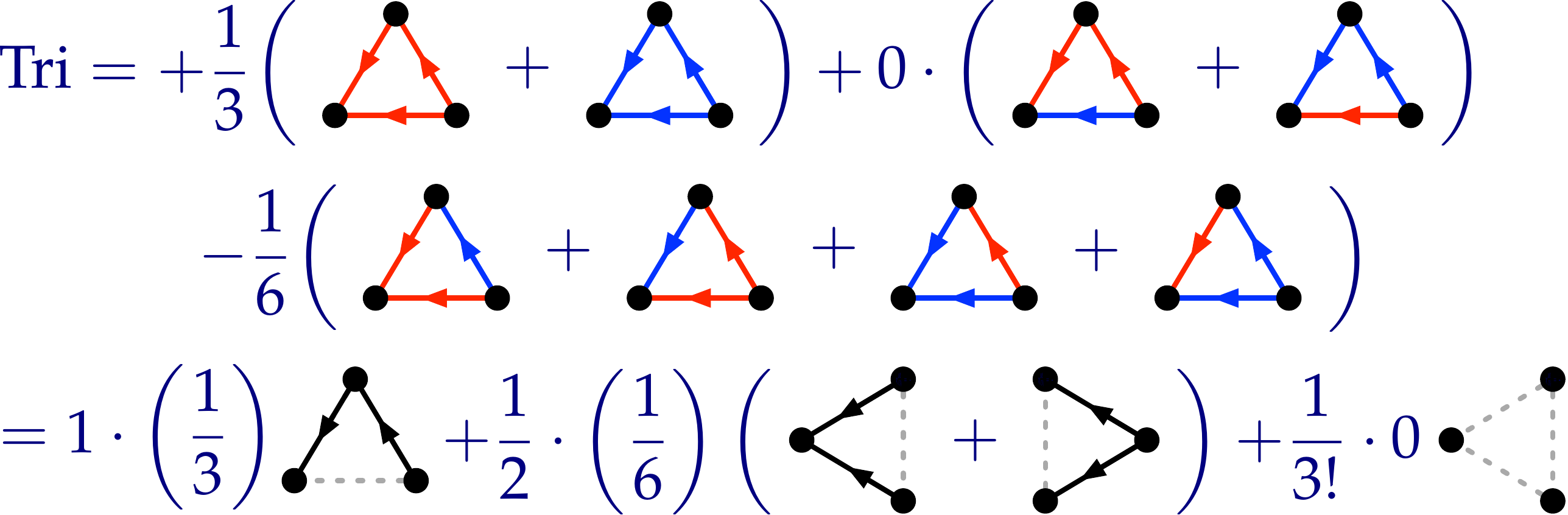}
    \caption{Color basis to BW basis: triangle.}
    \label{fig:trans-triangle}
\end{figure}

The next simplest example is the triangle. 
Fig.~\ref{fig:trans-triangle} shows the explicit map between the color basis and the BW basis. 
Based on the color parity, we exclude the odd-retarded, even-cut graphs from the outset. 
The 3-cut graph is not excluded by the color parity, but its $\omega$ value vanishes 
as a consequence of the zero-sum rule. 
It is the first instance of another general pattern: if a graph $G$ gets disconnected after all cut propagators are removed, then $\omega(G)=0$.

It is straightforward to verify the equivalence of the two bases in Fig.~\ref{fig:trans-triangle}. 
For the Wightman functions, it is a simple change of variables. 
Some care is needed when we deal with the time ordering. For instance, 
\begin{align}
    &
    (\W_A)_{12}\mem (\W_A)_{13}\mem (\W_S)_{23} 
    \,
    \Theta_{12}\mem \Theta_{13}\mem \Theta_{23}
    \\
    &\arrow
\begin{aligned}[t]
&
    (\W_A)_{12}\mem (\W_A)_{13}\mem (\W_S)_{23} 
    \,
    \Theta_{12}\mem \Theta_{13} 
    \,
    \tfrac{1}{2}
    \bigbig{
        \Theta_{23} +\Theta_{32}
    }
% \\
%     &
    = \tfrac{1}{2}\,
        (G_\text{ret})_{12}\mem
        (G_\text{ret})_{13}\mem
        (G_\text{cut})_{23} 
    \,.
\end{aligned}
\nonumber
\end{align}
In the first line, we used the symmetry between the vertex labels 2 and 3 inside the integral. The last step follows from the identity,  
$\Theta_{ij} + \Theta_{ji} = 1$. The resulting factor $1/2$ correctly captures the symmetry factor of the graph 
in the BW basis.

\begin{figure}[htbp]
    \centering
    \includegraphics[width=0.5\linewidth]{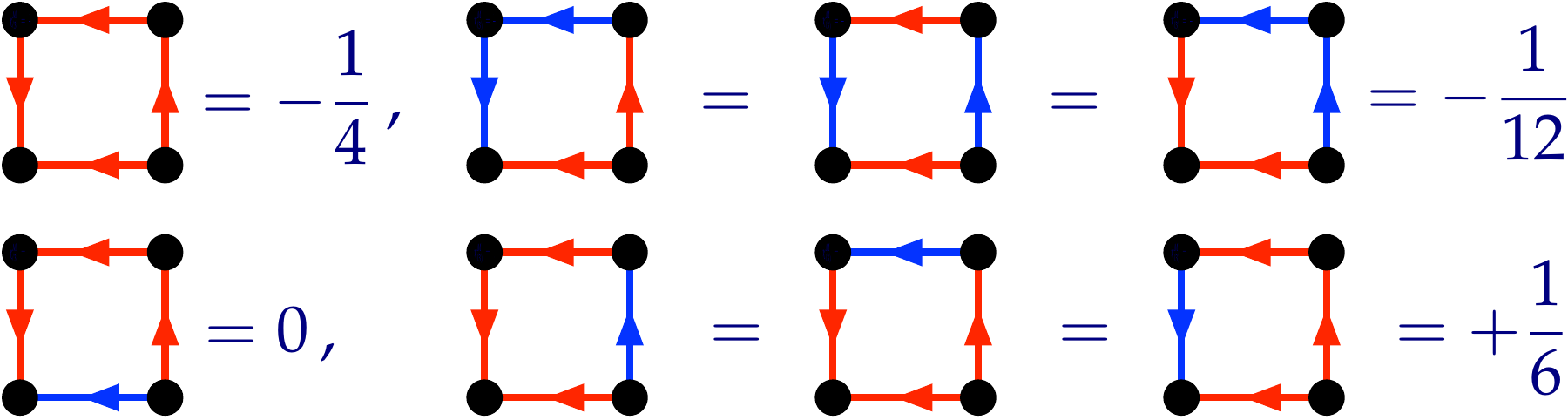}
    \caption{Colored trapezoids and their $\omega$ values.}
    \label{fig:bicolor-trapezoid}
\end{figure}

\begin{figure}[htbp]
    \centering
    \includegraphics[width=0.5\linewidth]{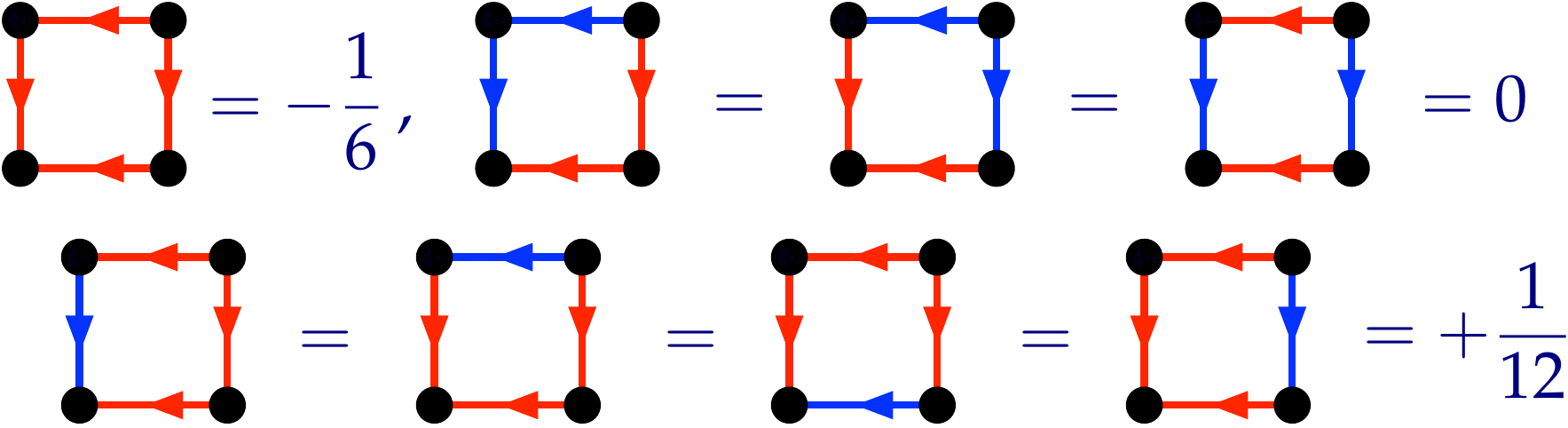}
    \caption{Colored diamonds and their $\omega$ values.}
    \label{fig:bicolor-diamond}
\end{figure}

\begin{figure}[htbp]
    \centering
    \includegraphics[width=0.5\linewidth]{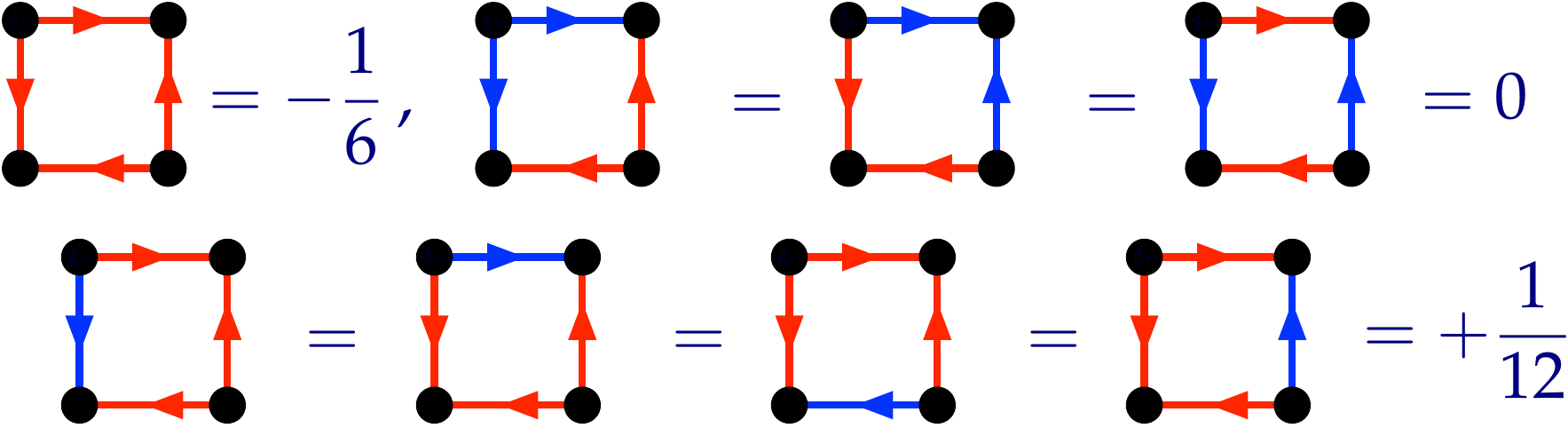}
    \caption{Colored folded diamonds and their $\omega$ values.}
    \label{fig:bicolor-dia-fold}
\end{figure}

\paragraph{4-point, Color} 
In Fig. \ref{fig:bicolor-trapezoid}, \ref{fig:bicolor-diamond}, and \ref{fig:bicolor-dia-fold}, we display one-loop quadrangle graphs in the color basis. 
To save space, we show only half of all possible colorings. 
The other half follows from the color parity.

\paragraph{Removable edges}

For single-colored graphs in the color basis, an edge whose time-ordering implication is redundant can be removed when computing $\omega$ values.
Examples include the bottom edge of the first two triangles in Fig.~\ref{fig:trans-triangle} and the bottom edge of the first quadrangle in Fig.~\ref{fig:bicolor-trapezoid}. 
We leave a detailed discussion of this phenomenon, and its generalization to mixed-color graphs, to a companion paper \cite{our-math-paper}.

%\newpage 
\paragraph{4-point, BW}

When a cut propagator is erased from a graph, $G$, we face two possibilities. 
If the reduced graph is disconnected, $\omega(G)=0$. 
If the reduced graph is connected, $\omega(G)$ is $\omega$ of the reduced graph. 

\begin{figure}[htbp]
    \centering
    \includegraphics[width=0.72\linewidth]{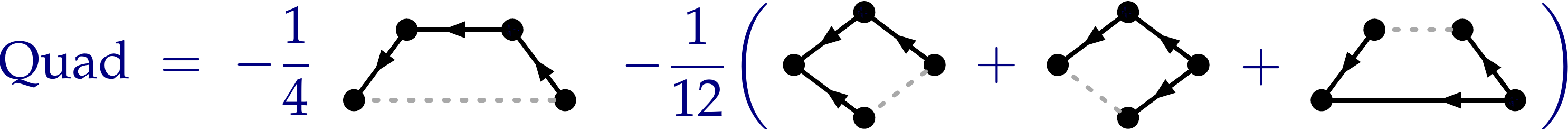}
    \caption{4-point, one-loop sum over graphs (non-zero $\omega$ only).}
    \label{fig:bw-1-loop-quad}
\end{figure}

If we apply these rules to one-loop graphs, 
together with the color parity rule in \eqref{color-parity-BW}, 
we reproduce the observation of \rcite{Pichini} that all one-loop graphs with non-vanishing $\omega(G)$ should contain precisely one cut propagator, 
and their $\omega$ values coincide with the tree ones. 
We have seen the one-loop bi-angle example in Fig.~\ref{fig:trans-biangle}, 
and the one-loop triangle example in Fig.~\ref{fig:trans-triangle}. 
A new example with quadrangles is given in Fig.~\ref{fig:bw-1-loop-quad}.

\begin{figure}[htbp]
    \centering
    \includegraphics[width=0.54\linewidth]{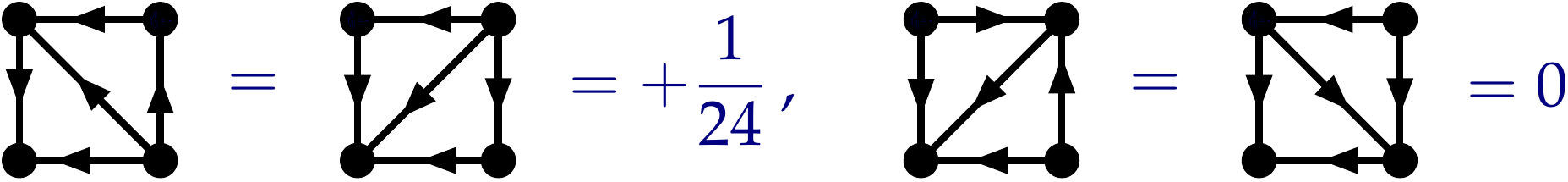}
    \caption{Some 4-point, 2-loop graphs and their $\omega$ values.}
    \label{fig:bw-2-loop}
\end{figure}

More interesting possibilities appear at higher loops. 
For 4-point (quadrangle) graphs, the color parity implies that the number of retarded propagators should be odd. Graphs with three retarded propagators inherit $\omega$ values from the corresponding tree graphs. 
Graphs with five retarded propagators are not directly related to tree graphs. Some examples are shown in Fig.~\ref{fig:bw-2-loop}. 

\newpage
%%%%%%%%%%%%%%%%%%%%%%%%%%%%%%%%%%%%%%%%%%%%%%%%%%%%%%%%%%%%%%%%%%%%%
\section{From Magnus to Murua}
\label{sec:Murua}

The $n$-th order Magnus recursion in \eqref{Magnus-recursion} involves level $k$ nested commutators with $1 \le k \le n-1$. 
For small $n$, say up to $n=4$, it is easy to evaluate the commutators by hand. 
But the computation becomes cumbersome quickly as $k$ increases. 
It is desirable to find a shortcut for determining the values of $\omega(G)$ in the diagrammatic expansion in \eqref{Mag-schematic}. 

For rooted trees, a pioneering work by Murua~\cite{Murua} provided a recursive formula to compute $\omega(G)$. The formula was extended to non-rooted trees in \rcite{KKKL}. 
Our goal in this section is to further generalize the formula to cover all loop graphs, 
both in the color basis and in the BW basis. 
As a preparation, we give a brief review of the tree Murua formula of \rcite{KKKL}.

The extended Murua formula for a tree graph is given by
\begin{align}
\label{C-Murua}
    \omega(\tau)
    \,=
    \sum_{p \in P_s(\tau)}
        (-1)^{\ell(p')} B_{|p|-1}\,
        e(p')\,
        |p|\,
        \omega(\tau \backslash p) 
    \,.
\end{align} 
Here, $B_k$ are the Bernoulli numbers originating from the Magnus recursion. 
This equation can be viewed as a diagrammatic representation of \eqref{Magnus-recursion-QM}, where $(-1)^{\ell(p')} e(p') |p|$ encodes the nested commutator structure and $\omega(\t \backslash p)$ encodes $\chi_{r}$. 

For a rooted tree, the vertex $s$ refers to the unique root, which is the ``global minimum" 
according to the ordering of vertices set by the directed edges. 
For a non-rooted tree, the vertex $s$ can be one of the semi-roots, a ``local minimum.'' 
The formula involves the choice of a semi-root $s$, but the resulting $\omega(\tau)$ turns out to be independent of the choice. $P_s(\tau)$ is the set of partitions of $\tau$ that contain all edges connected to the semi-root $s$.

A partition $p$ of a tree $\tau$ is an arbitrary collection of edges of $\tau$; there are $2^{E(\tau)}$
partitions when there is no restriction. It is understood that $p$ is made connected by
shrinking each connected component of the remainder $(\tau\backslash p)$. 
 Given a partition $p$, we flip the orientation of all the ``wrong"
edges of $p$ to obtain a new tree $p'$ rooted with respect to $s$.  The function $\ell(p')$ counts how many edges of $p$ should be flipped to reach $p'$. 

For a tree graph, the function $e(\tau)$ is defined as 
\begin{align}
\label{e-def-tree}
    e(\tau) \,=\, 
        \frac{1}{|\tau|!}\, \phi(\tau)
    \,,
\end{align}
where the integer-valued $\phi(\tau)$ counts the number of linear extensions of a poset. 
In other words, it counts the number of all possible orderings of labeled vertices that are
consistent with the directed edges of a tree. 
See Fig.~5 of \rcite{KKKL} for some examples. 

The definition of the $e$-function generalizes straightforwardly to any acyclic loop graph. 
For a cyclic graph, where a fully consistent ordering of vertices is impossible, 
it is convenient to set $e=0$. 
The definition of the $e$ function implies the contraction rule:
\begin{align}
\label{e-contraction}
    e(G_{[1 \leftarrow 2]}) \,+\, e(G_{[1 \rightarrow 2]}) 
    \,=\,
        e(G_{[1\vert 2]})
    \,.
\end{align}
On the LHS, we consider two graphs which differ only by the direction of the edge connecting a specified pair of vertices (labeled 1 and 2). On the RHS, the edge [1-2] is cut while the rest of the graph remains the same. 

The $e$ function can be further generalized to incorporate colors or undirected edges, 
but the one for fully directed single-color graphs will be sufficient in this paper.

%\newpage 
\subsection{Quantum Murua Formula in the Color Basis}

Our proposal for the quantum Murua formula in the color basis is as follows. 
\begin{align}
\label{Q-Murua}
    \omega(G) 
    \,= \sum_{p \in P_s(G)} \sum_{p'' \in F_s(p)} 
        (-1)^{\ell(p'')}\,  
        e(p'')\, B_{|p|-1} |p|\, \omega(G\backslash p) 
    \,.
\end{align}
As in the tree formula, the $(-1)^{\ell(p'')} e(p'') |p|$ factor reflects the nested commutator structure 
while $\omega(G \backslash p)$ encodes $\chi_{r}$ in \eqref{Magnus-recursion-QM}. 
Here, $s$ is a choice of semi-root and $P_s(G)$ is the set of partitions of $G$ that contain the semi-root. 
Again, the value of $\omega(G)$ is independent of the choice; 
otherwise Lorentz invariance of the Magnusian will be lost. 
The modified partition $p''$ is obtained from $p$ following a few steps, to be explained shortly.  
A critical complication compared to the tree case is that, for a given $p$, there can be multiple possibilities for $p''$ and we have to sum over all of them. 

\paragraph{Examples}

\begin{figure}[htbp]
    \centering
    \includegraphics[width=0.8\linewidth]{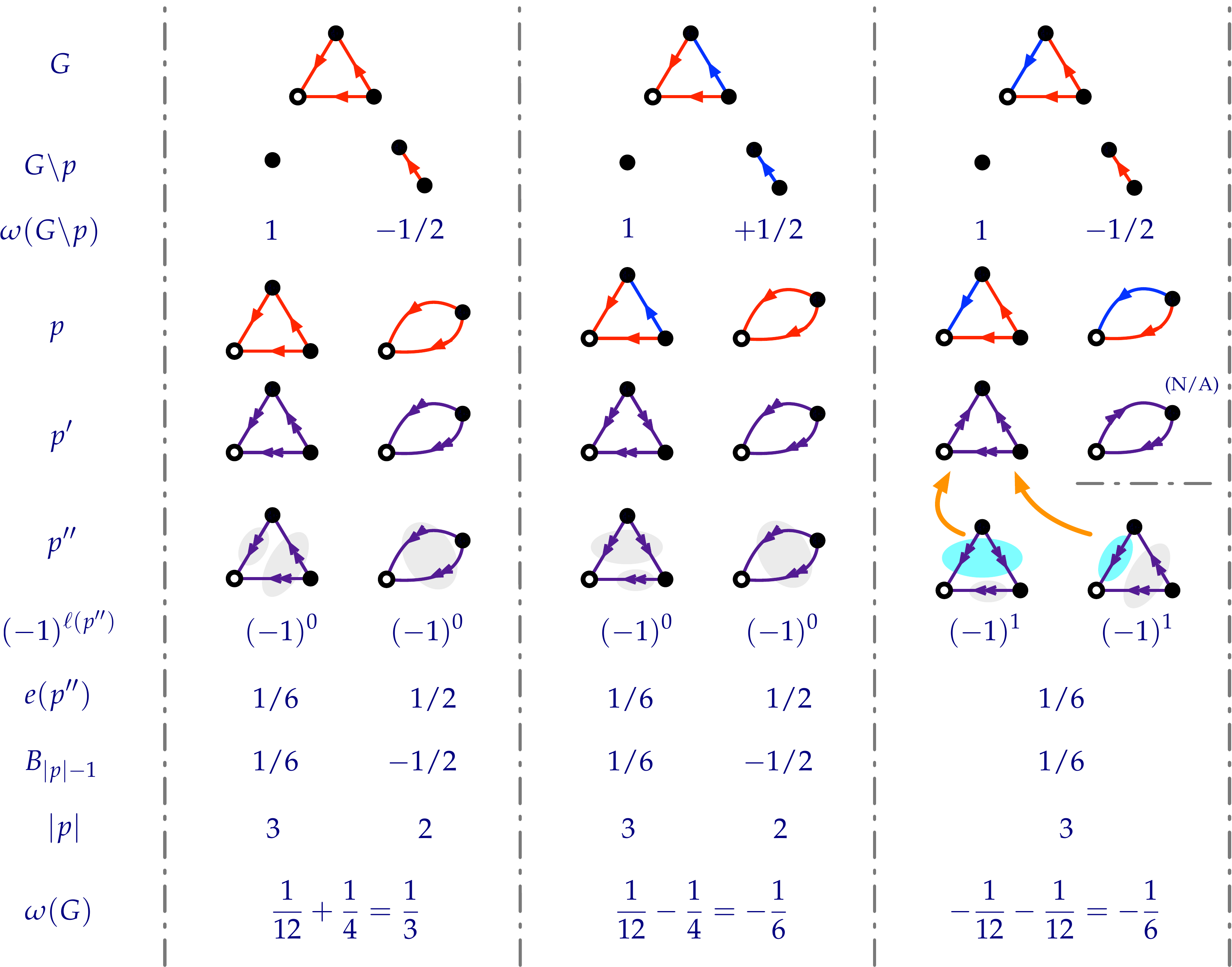}
    \caption{Loop-extended Murua formula for triangle graphs.}
    \label{fig:MS-triangle}
\end{figure}

Before discussing the formula in general, we present a few concrete examples. Fig.~\ref{fig:MS-triangle} shows the formula applied to triangle graphs. 
Note that the $\omega$ values of the second column and the third column are the same, as the two graphs are related by the overall time-reversal, but the Murua formula operates differently. 

\begin{figure}[htbp]
    \centering
    \includegraphics[width=0.95\linewidth]{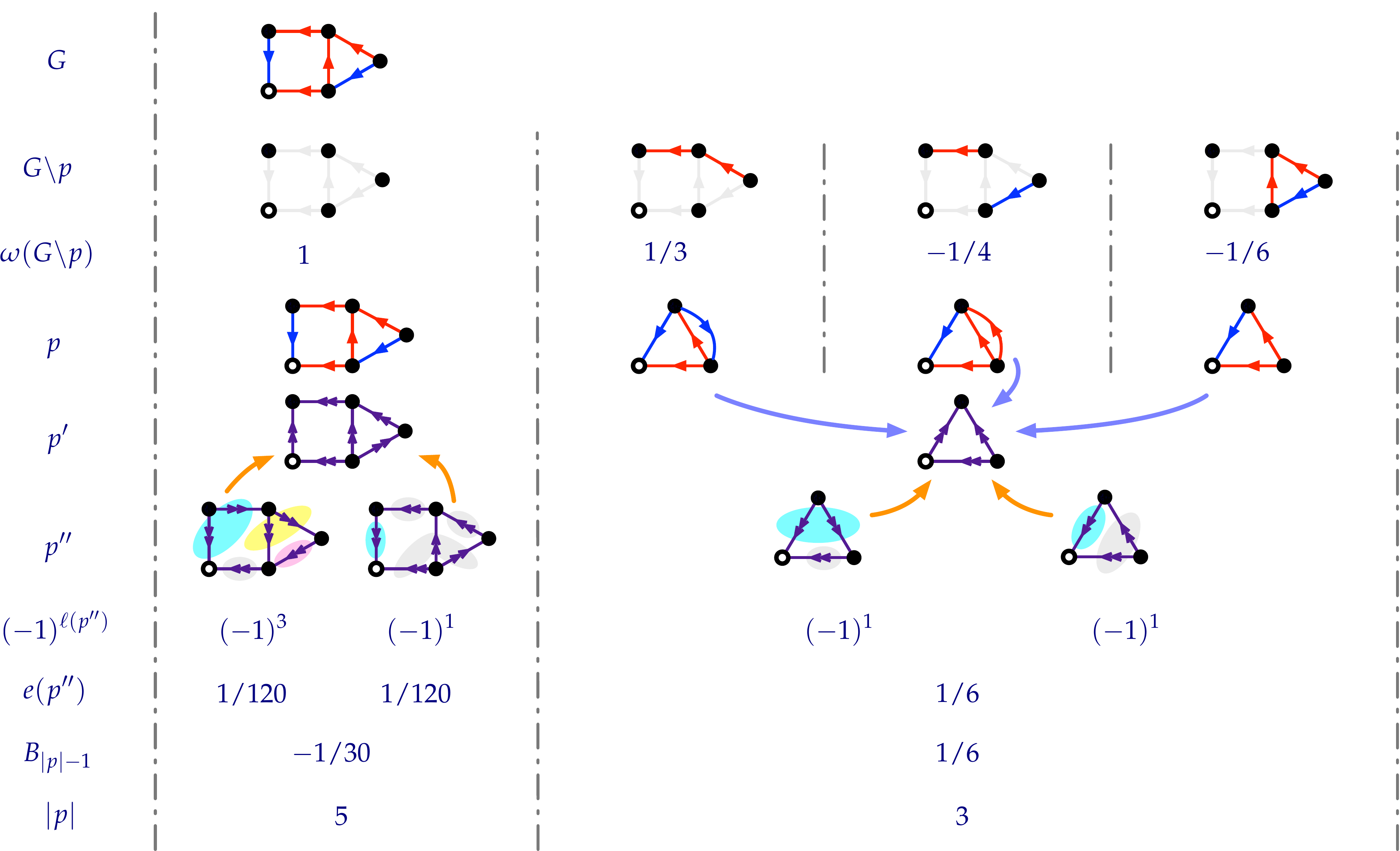}
    \caption{Loop-extended Murua formula for a pentagon graph.}
    \label{fig:MS-penta}
\end{figure}

%\newpage 
Fig.~\ref{fig:MS-penta} shows the formula applied to a pentagon with a diagonal inside. It hosts $2^4=16$ partitions, but we omit the 12 partitions which contribute zero to $\omega(G)$. 
The net result of the formula can be summarized as 
\begin{align}
\begin{split}
  \omega(G) &\,=\,
    1\cdot\left( \frac{(-1)^3}{120} + \frac{(-1)^1}{120} \right) \left(-\frac{1}{30}\right)\cdot 5 + \left( \frac{1}{3} - \frac{1}{4} -\frac{1}{6} \right)\cdot\left( \frac{(-1)^1}{6} \cdot 2\right)\cdot \frac{1}{6} \cdot{3}  
    \\
    &\,=\, \frac{1}{360} + \frac{1}{72} = \frac{1}{60} \,. 
\end{split}
\end{align}

\paragraph{First glance at the formula} 

We assume that the reader has become familiar with the classical Murua formula in \eqref{C-Murua}, and give a step-by-step guide on how to use the quantum Murua formula in \eqref{Q-Murua}. 

We choose a semi-root and list all partitions containing the semi-root. The decomposition of $G$ into $p$ and $G\backslash p$ proceeds as usual, whether $G$ contains loops or not. The factors, $|p|$, $B_{|p|-1}$, $\omega(G\backslash p)$ contribute in the same way as in the tree case. 
We remind the reader that $|p|$ is the number of \emph{vertices} of the partition, after $G\backslash p$ has been shrunken to points. 

The challenge with the quantum Murua formula is how to turn $p$ into the set of $p''$'s. 
The first step is to forget the time ordering and focus on the operator ordering. In practice, we reverse the direction of all blue edges and then ignore all colors. If a cycle is formed, we discard the partition and proceed to the next one. If an acyclic banana loop is formed, we replace it with a simple edge. The outcome is called $p'$. 
In the sense discussed in \Sec{sec:commutators}, $p'$ pays attention to the operator ordering and ignores the time ordering. 
The doubled arrowheads in $p'$ of Figs.~\ref{fig:MS-triangle}-\ref{fig:MS-penta} reflect the shift of focus to operator ordering.

The second step is to look up the operator product diagrams 
discussed in \Sec{sec:commutators}, 
in order to enumerate all possible candidates $p''$ 
which can match $p'$ upon systematic flipping of directions. 
To explain how it works, we should digress to take a closer look at the operator product diagrams. 

%\newpage 
\paragraph{Almost-rooted loops}

Recall the operator products with exponential Wick operators in Fig.~\ref{fig:spider-2}. 
Expanding the exponential and extracting \emph{connected} primary graphs, we obtain a set of graphs at each level. 
Let us call the results ``spider webs." 

\begin{figure}[htbp]
    \centering
    \includegraphics[width=0.6\linewidth]{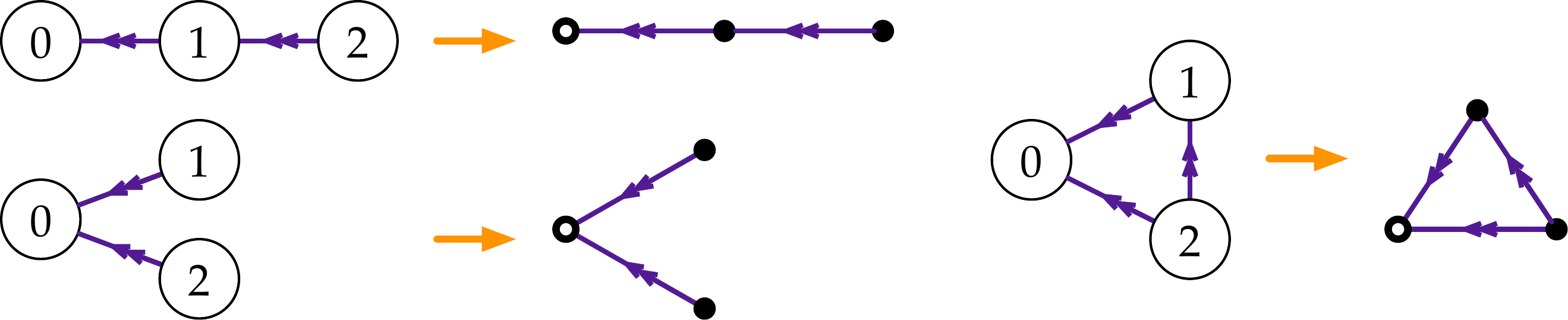}
    \caption{Spider webs at level 2.}
    \label{fig:popping-2}
\end{figure}

\begin{figure}[htbp]
    \centering
    \includegraphics[width=0.56\linewidth]{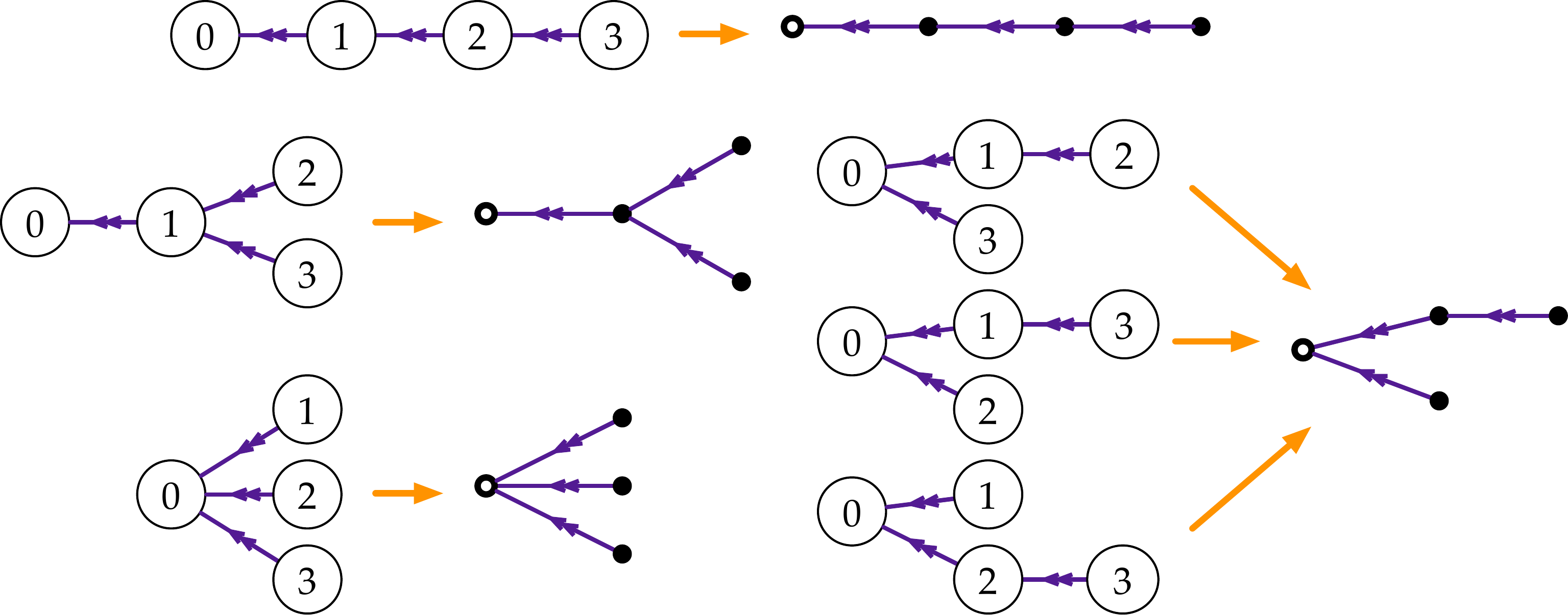}
    \includegraphics[width=0.56\linewidth]{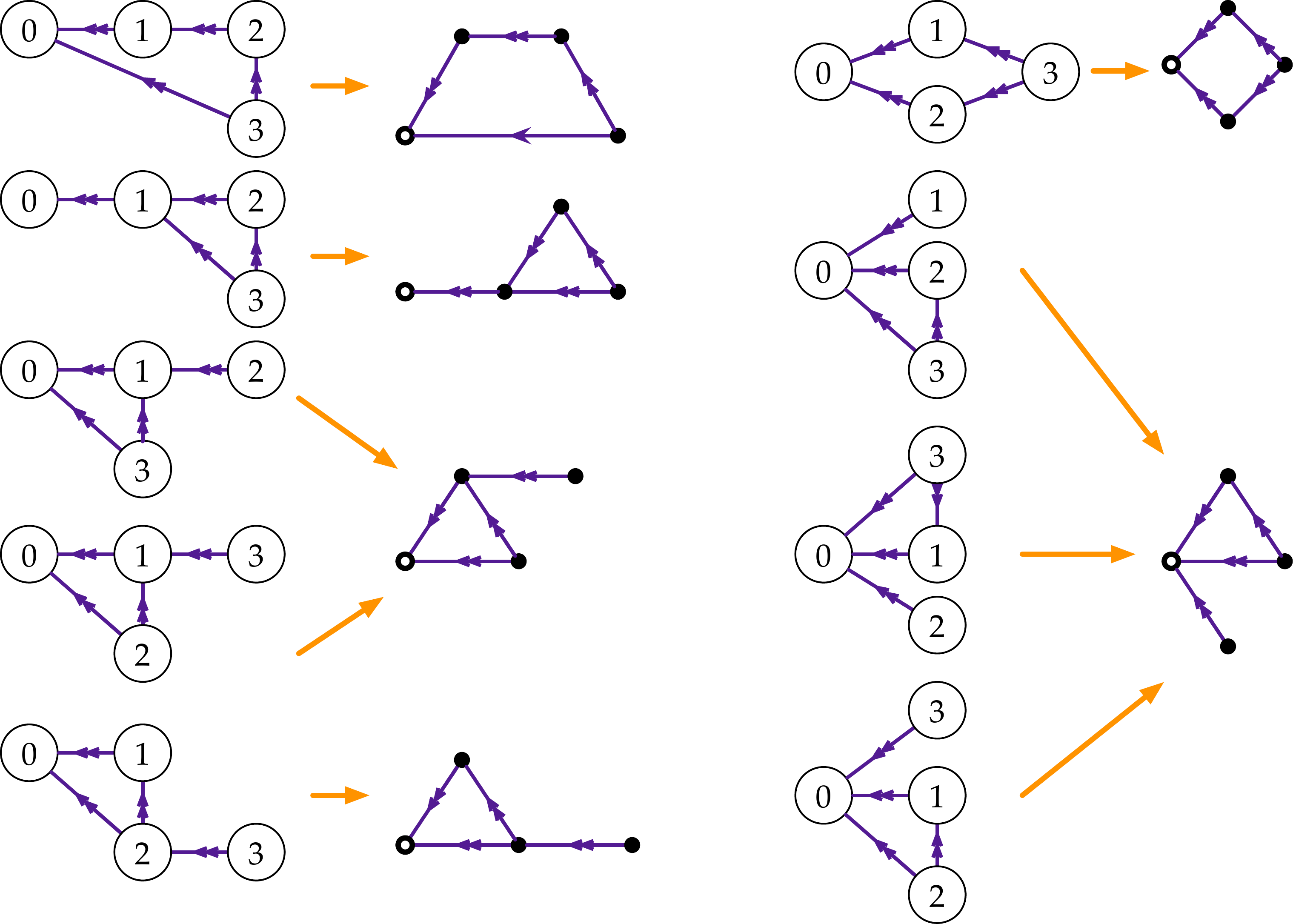}
    \includegraphics[width=0.7\linewidth]{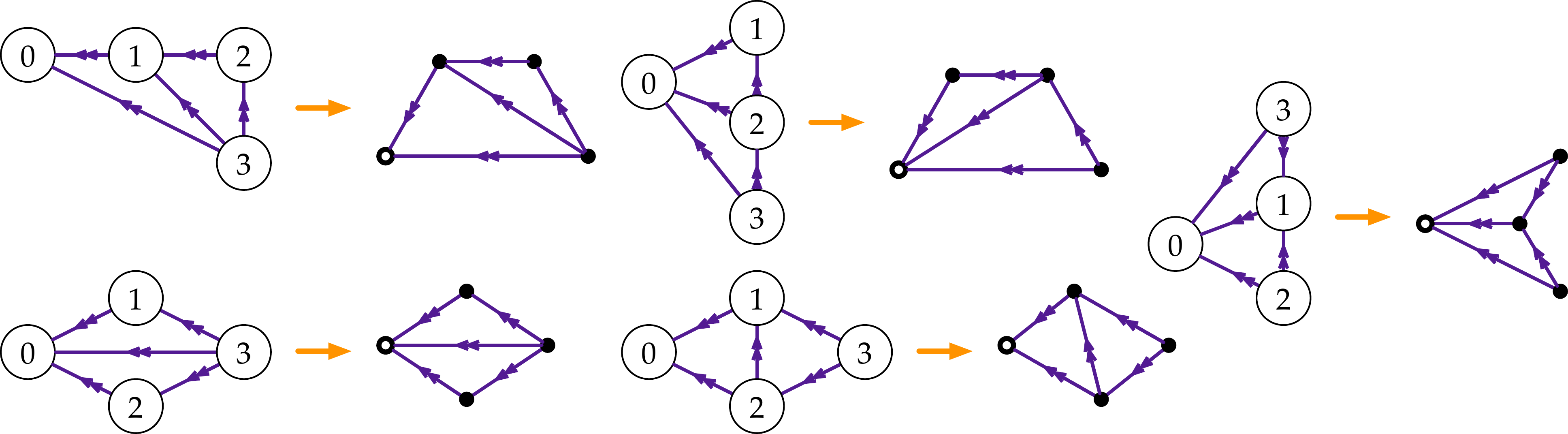} \\
    \includegraphics[width=0.26\linewidth]{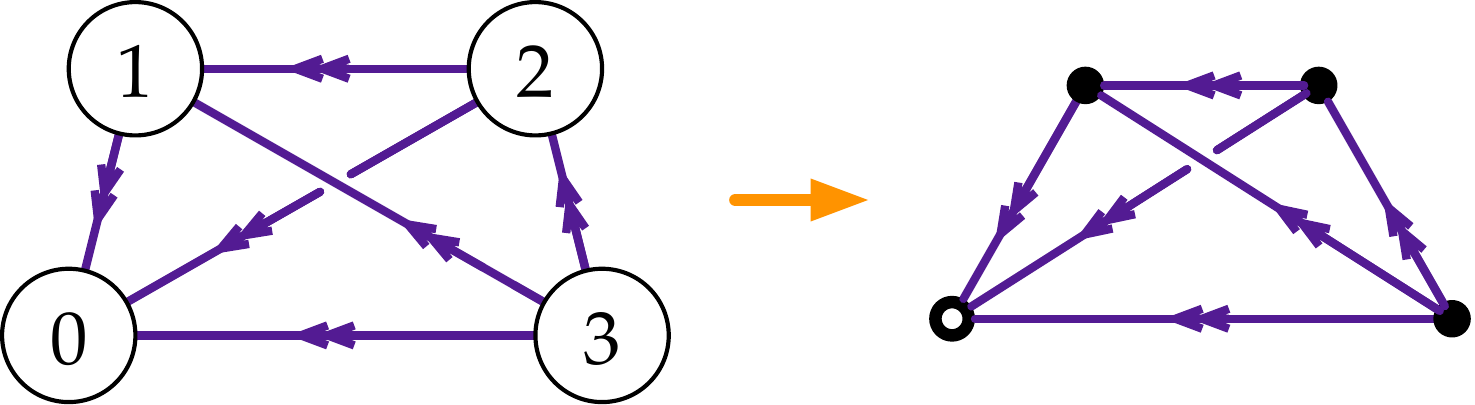}
    \caption{Spider webs at level 3.}
    \label{fig:popping-3}
\end{figure}

In the first step of spider web action, we keep all the vertex labels, but in the second step we ``forget" the vertex labels except the root vertex $\mathcal{V}_0$. 
The procedure applied to level $2,3$ is shown in Fig.~\ref{fig:popping-2} and Fig.~\ref{fig:popping-3}, respectively. 

We emphasize again that the directed edges in the figures signify the operator ordering and not the time ordering, as indicated by the doubled arrowheads. As noted in \rcite{PSFOR-S}, no cyclic graph can appear. 
Starting from $k=3$, we find cases where two or more vertex-labeled graphs end up with the same reduced graph: a phenomenon observed for trees in \rcite{KKKL} now being generalized to the loops. 
A nice feature of the Murua formula is that the combinatorics of this many-to-one map is automatically incorporated by the $e(p'')$ factor.

\begin{figure}[htbp]
    \centering
    \includegraphics[width=0.64\linewidth]{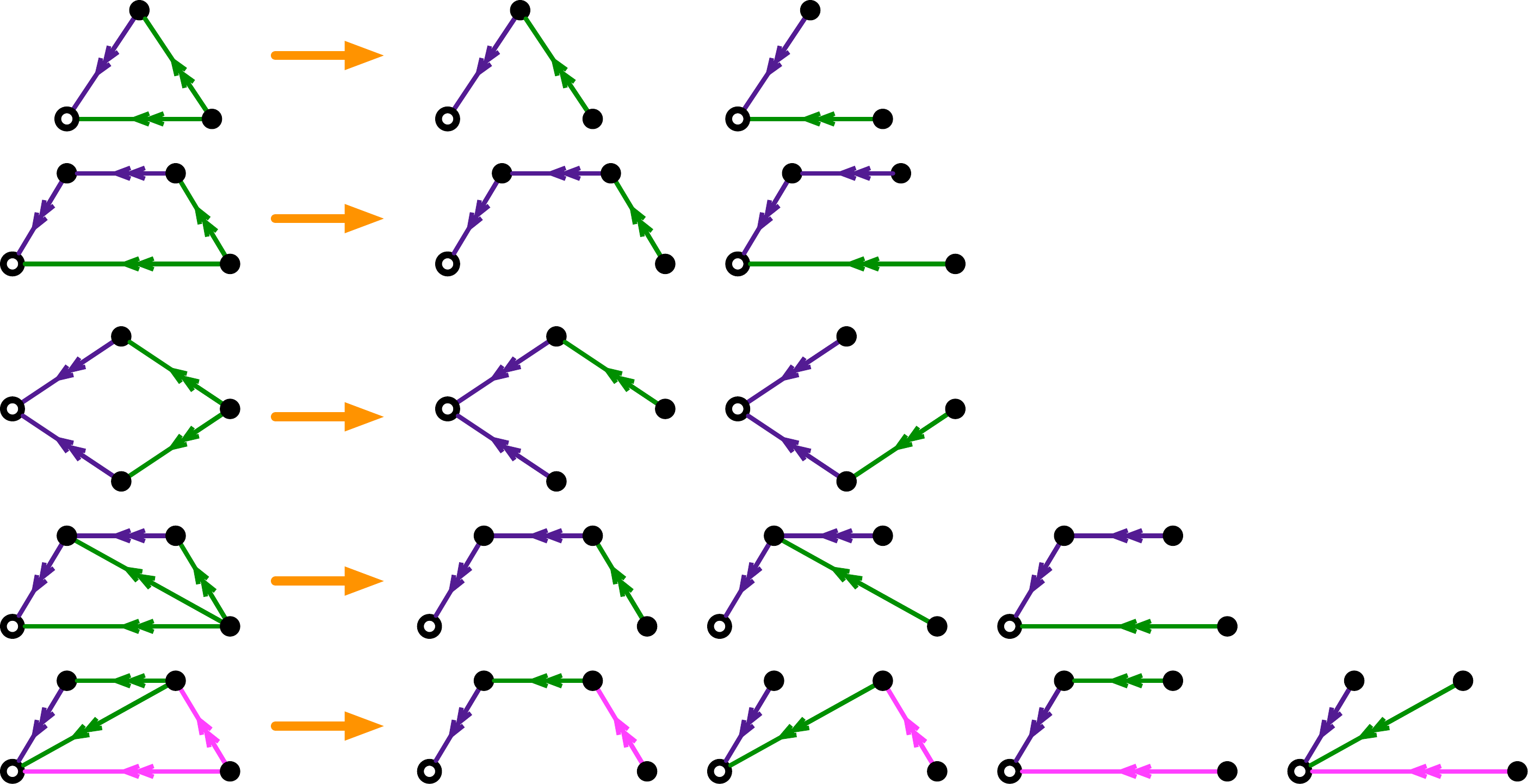}
    \caption{Almost-rooted loop graphs. The tree sub-graphs resulting from the procedure explained in the text are always rooted.}
    \label{fig:almost-root}
\end{figure}

It is clear from the figures that all trees produced by this procedure are rooted. 
A similar observation holds for loop graphs. 
Each vertex in the reduced graph, except the root, carries at least one outgoing edge. Let $o_j$ be the number of outgoing edges of vertex $j$. 
Given a reduced loop graph, we generate trees by ``cutting" some edges under a set of conditions. 
For vertices with $o_j =1$, we do nothing. For vertices with $o_j > 1$, we cut all but one edge. Which edges to cut is a matter of choice. We may consider all possibilities. As shown in Fig.~\ref{fig:almost-root}, the result of the cutting is always a connected, rooted tree, no matter which choice we make. 
In this specific sense, we will say that the loops in Fig.~\ref{fig:popping-2} and Fig.~\ref{fig:popping-3} are ``almost-rooted".

\paragraph{Back to the quantum Murua formula} 

We are now ready to explain the most subtle step of the quantum Murua formula, namely, figuring out the set of $p''$ for a given $p'$. Generically, $p'$ is not almost-rooted. The candidates for $p''$ are taken from the spider web graphs, so by construction $p''$ is almost-rooted. The level of spider web is fixed to be $V(p)-1 = E(p)-L(p)$, but there can be multiple $(p'')$'s that match the same $p'$. We should find each and every almost-rooted $p''$ which can reach $p'$ by a sequence of flips of directed edges. 

On the $p''$ side, we have a clear distinction between incoming and outgoing edges. To map it to $p'$, for each vertex of $p''$, we may choose to flip all outgoing arrows simultaneously or not to flip them at all; partial flipping ({\it e.g.}, flipping 2 out of 3 edges from the same vertex) is strictly forbidden. 
The sign factor $(-1)^{\ell(p'')}$ in the Murua formula counts the number of vertices on which we flip the outgoing arrows to map $p''$ to $p'$.

The sign flip is based on the relative signs among various operator products coming from a common nested commutator in \eqref{commutator-expanded}. A crucial point in the derivation of the Murua formula is that the time ordering among non-root vertices in the Magnus recursion relation in \eqref{Magnus-recursion-QM} is not fixed, so we have to consider all permutations of non-root vertices. 
For example, in the right-most column of Fig.~\ref{fig:MS-triangle}, the two cases of $(p'')$ come from two different time orderings of the level 2 nested propagator.

%\newpage 
\paragraph{Contraction rules} 

It is possible to implement the quantum Murua formula on a computer, 
but it tends to be slow as $n$ increases. 
As in the tree case~\cite{KKKL}, 
a more efficient route is to derive the ``edge-contraction" rules (not to be confused with the Wick contraction of QFT) from the Murua formula and then use the rules to compute $\omega(G)$. 

In the color basis, the edge-contraction rules can be stated as 
\begin{subequations}
\label{contraction-color}
\begin{align}
    \omega(G_{\textcolor{red}{[1\leftarrow 2]}}) -\omega(G_{\textcolor{blue}{[1 \rightarrow 2]}})
    \,&=\, 
        -\omega(G_{[1\cdot 2]}) \,,
    \label{contraction-color-a}
    \\
    \omega(G_{\textcolor{red}{[1\leftarrow 2]}}) +\omega(G_{\textcolor{blue}{[1 \leftarrow 2]}})
    \,&=\,
        \omega(G_{[1 \vert 2]\text{conn}})
    \,. 
    \label{contraction-color-b}
\end{align}
\end{subequations}
Taking linear combinations of the two rules, we obtain contraction rules considering one color at a time, 
\begin{subequations}
\label{contraction-color-comb}
    \begin{align}
    \omega(G_{\textcolor{red}{[1\leftarrow 2]}}) +\omega(G_{\textcolor{red}{[1 \rightarrow 2]}}) 
    &\,=\, 
        - \omega(G_{[1\cdot 2]}) + \omega(G_{[1 \vert 2]\text{conn}}) 
    \,,
    \label{contraction-color-c}
\\
    \omega(G_{\textcolor{blue}{[1\leftarrow 2]}}) +\omega(G_{\textcolor{blue}{[1 \rightarrow 2]}}) 
    &\,=\,
        + \omega(G_{[1\cdot 2]}) + \omega(G_{[1 \vert 2]\text{conn}}) 
    \,.
    \label{contraction-color-d}
\end{align}
\end{subequations}

\begin{figure}[htbp]
    \centering
    \includegraphics[width=0.5\linewidth]{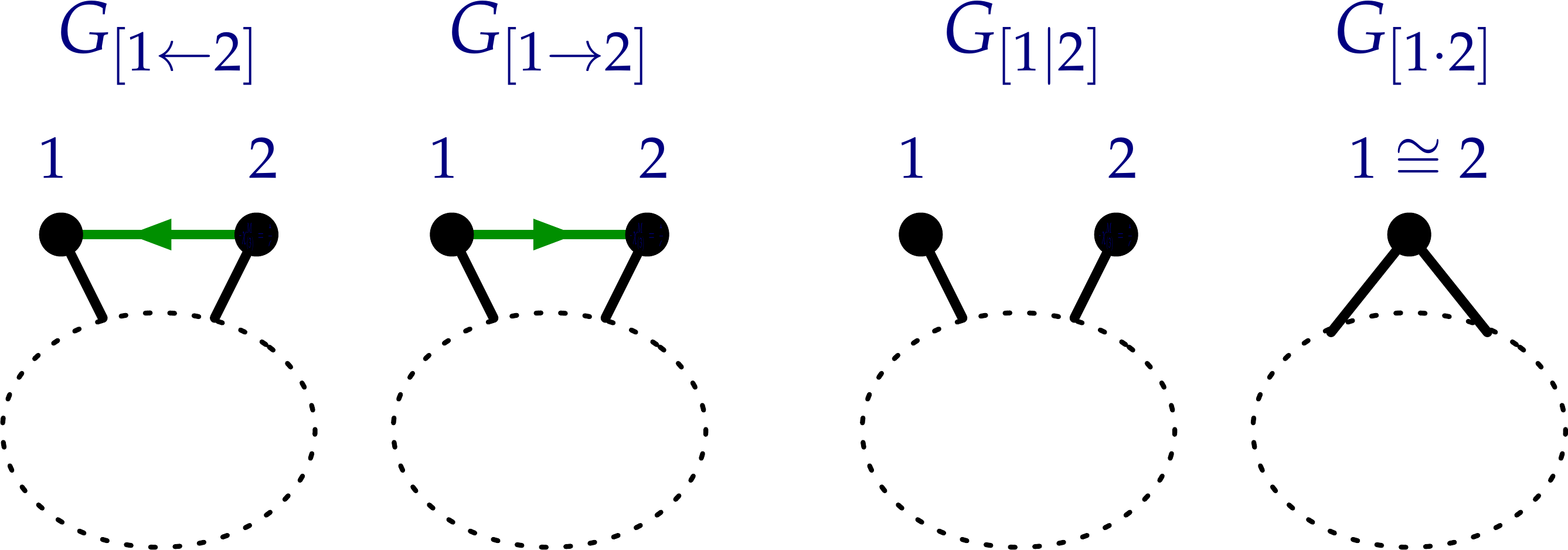}
    \caption{The configurations for contraction rules. The color can be either red or blue.}
    \label{fig:contraction-monochrome}
\end{figure}

On the LHS of the contraction rules, we consider both orientations of the edge connecting vertices 1 and 2, as well as both colors. The rest of the graph is assumed to be identical. On the RHS, $G_{[1|2]}$ means the edge has been cut. For trees $G_{[1|2]}$ would be a formal product of two disjoint trees (called forest). Since the removable edge is a part of a loop, $G_{[1|2]}$ in the current context remains connected. The notation $G_{[1|2]\text{conn}}$  means that the term should be included only when $G_{[1|2]}$ is connected. Otherwise, it will spoil the tree-level contraction rule. Finally, $G_{[1\cdot 2]}$ means that the edge has collapsed to merge the two vertices. 
See Fig.~\ref{fig:contraction-monochrome} for an illustration.

We wish to prove the two rules, \eqref{contraction-color-a} and \eqref{contraction-color-b}. 
The proof proceeds recursively. The rules hold for $E(G)=1$. Then we assume that the rule holds for $E(G) \le r \in \mathbb{Z}_{\ge 1}$ and try to prove the same rule for $E(G) = r+1$.

\eqref{contraction-color-a} is easier to prove.  
When the edge [1-2] belongs to the partition $p$, the red case and the blue case give the same $p'$, which cancel each other. These are precisely the partitions that appear only on the LHS and not on the RHS. 
When the edge [1-2] falls into $G\backslash p$, we can apply the same rule for smaller graphs to prove \eqref{contraction-color-a} for the whole graph. 

\begin{figure}[htbp]
    \centering
    \includegraphics[width=0.95\linewidth]{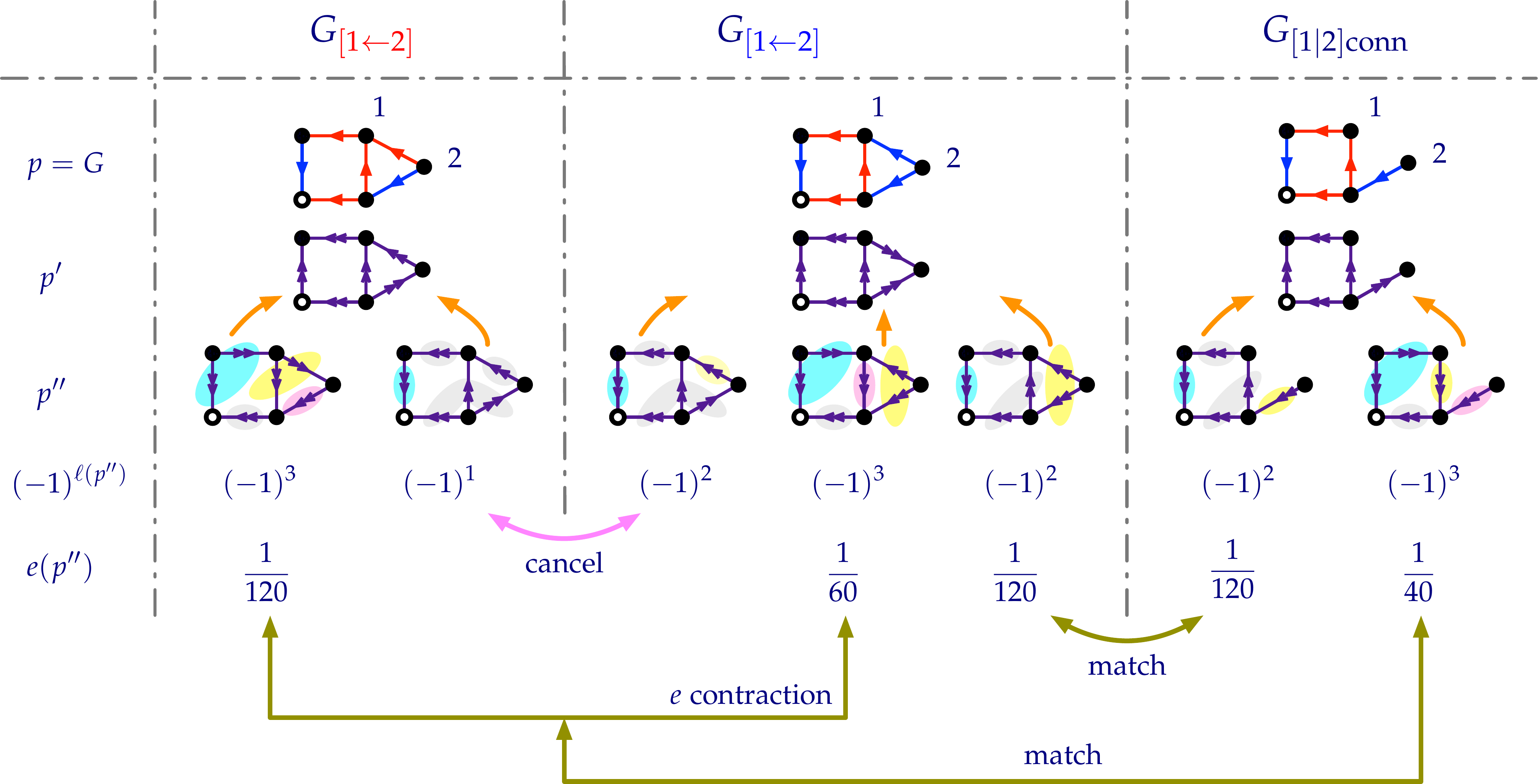}
    \caption{Cancellation and matching of $p''$.}
    \label{fig:MS-contraction}
\end{figure}

The proof of \eqref{contraction-color-b} is more involved. 
When the edge [1-2] is included in $G\backslash p$, we can again apply the same rule for smaller graphs, 
so the recursive proof works trivially.
When the edge [1-2] belongs to $p$, the set of $(p'')$'s differs non-trivially 
between the two sides of \eqref{contraction-color-b}. 
The $(p'')$'s of the two graphs on the LHS 
can be divided into two groups:
``match'' or ``cancel.'' 
The $(p'')$'s in the first group pair up to match the corresponding $(p'')$ of the RHS. 
The value of $e(p'')$ matches via the $e$-contraction rule in \eqref{e-contraction}. 
When any $p''$ is cyclic we take $e(p'')=0$. 
Those $(p'')$'s in the second group cancel pairwise. 
The canceling pair of $(p'')$'s on the LHS are identical between the red and blue parent graphs
but carry the opposite $(-1)^{\ell(p'')}$. 
An example illustrating the idea of ``match or cancel" is given in Fig.~\ref{fig:MS-contraction}. 

The contraction rule is strong enough to compute $\omega(G)$ for all graphs recursively. Our starting point is the set of $\omega$ values for tree graphs, $\omega(\tau)$, previously obtained in \rcite{KKKL}. For the colored trees, 
$\omega(\tau)$ is modulated by $(-1)^\text{blue}$. 
The recursive construction works if one can relate $\omega(G)$ of an $L$-loop graph to those of $(L-1)$-loop graphs. 

Given a non-colored directed graph with $E$ edges, we consider the family of $2^{E}$ colored graphs all at once.
\eqref{contraction-color-b} relates $\omega$ values of two graphs which differ only by the color of an edge, up to $\omega$ of an $(L-1)$-loop graph. Clearly, all $2^{E}$ members of the colored family are related this way. 
If we know the $\omega$ value of at least one member of the family in advance, the $\omega$ values of all the other members of the family are determined by the contraction rule. 
Fortunately, as pointed out in \rcite{PSFOR-S}, a graph carrying a cycle in the sense of operator ordering has $\omega =0$. 
It is easy to show that any family of colored loop graphs contains at least one such graph. 

\begin{figure}[htbp]
    \centering
    \includegraphics[width=0.6\linewidth]{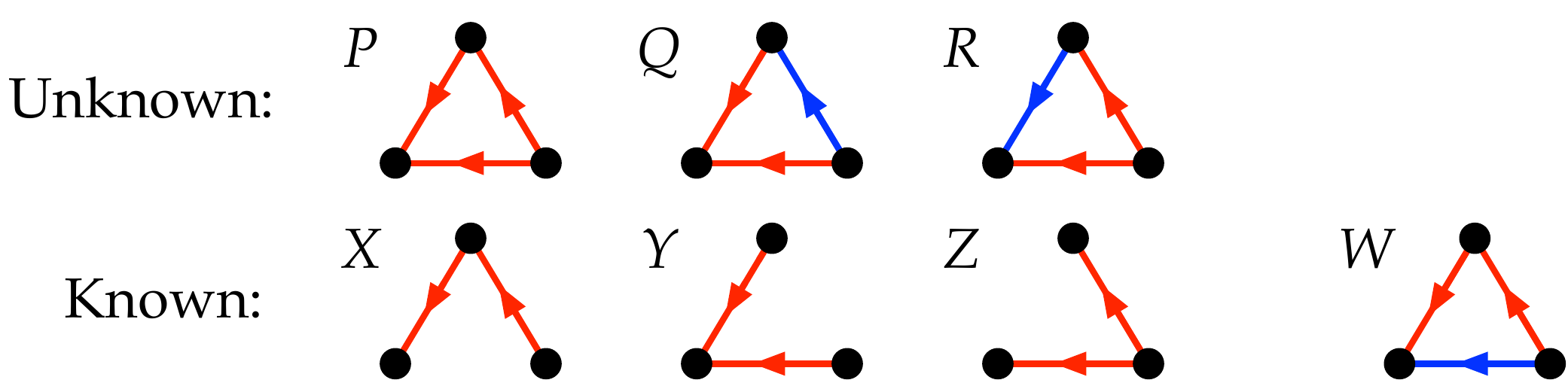}
    \caption{Contraction rule applied to one-loop graphs.}
    \label{fig:contraction-color}
\end{figure}

Fig.~\ref{fig:contraction-color} illustrates how to determine the $\omega$ values of one-loop triangle graphs using the contraction rule, $\omega$ for tree graphs, and a known loop graph with $\omega=0$. 
Our goal here is to compute the $\omega$ values of the one-loop graphs ($P, Q, R$) using the known values of $\omega$ for the other four graphs. The tree values are $\omega(X)=1/3$, $\omega(Y)=\omega(Z)=1/6$. We recognize graph $W$ as one with $\omega=0$. 
The contraction rule, \eqref{contraction-color-b}, implies 
\begin{align}
    \begin{split}
        \omega(P) + \omega(W) \,&=\, \omega(X) \,,
        \\
        \omega(P) + \omega(Q) \,&=\, \omega(Y) \,,
        \\
        \omega(P) + \omega(R) \,&=\, \omega(Z) \,.
    \end{split}
\end{align}
The solution, $\omega(P) = 1/3$, $\omega(Q) = \omega(R) = -1/6$, 
agrees with \rrcite{Pichini,PSFOR-S} 
and Fig.~\ref{fig:trans-triangle}. 

\paragraph{Zero-sum rule} 

Given a directed, acyclic, non-colored graph with $E$ edges, 
we observe that the sum of $\omega(G)$ over $2^E$ colored graphs sharing the same non-colored graph vanishes. 
Let us call it a ``zero-sum rule.'' 
It is easy to prove recursively using \eqref{contraction-color-b}. 
The reader is encouraged to verify the rule in simple examples in Figs.~\ref{fig:trans-biangle} and \ref{fig:trans-triangle}. 

Physically, the zero-sum rule is a necessary condition for Lorentz covariance.
If the sum is nonzero, we can always find a configuration of mutually space-like separated points where its corresponding integrand $\sum \omega(G)\mem I(G)$ is nonzero,\footnote{$I(G)$ was used in place of $\mathcal{I}(G)$ to indicate that the expression refers to the integrand rather than the integral.} 
because there always exists an ordering of vertices that can satisfy the time-ordering relations imposed by the product of step functions. 
For such a configuration of spacetime points, there always exists a Lorentz transformation that maps it to a different configuration where the time-ordering relations are violated, resulting in $\sum \omega (G) \mem I(G) = 0$.

%\newpage 
%%%%%%%%%%%%%%%%%%%%%%%%%%%%%%%%%%%%%%%%%%%%%%%%
\subsection{Quantum Murua Formula in the BW Basis}

The quantum Murua formula in the BW basis can be stated as follows.
\begin{align}
\label{Q-Murua-BW}
    \omega(G) \,= \sum_{p \in P_s(G)} \sum_{p'' \in F_s(p)}\, (-1/4)^{m(p'')} (-1)^{\ell(p'')}\,   e(p'')\, B_{|p|-1} |p| \,\omega(G\backslash p) \,.
\end{align}
The undirected edges play no role, and it suffices to consider graphs made of directed edges only. 
Partitions are taken in the usual way. The partitions with tadpoles are discarded. 
Unlike in the color basis, the partition $p$ may contain (non-tadpole) cycles. 
The cycles do \emph{not} mean a contradictory time ordering such as $\Theta_{12} \Theta_{23} \Theta_{13}$ 
since the vertices of $p$, being connected components of $G\backslash p$, 
can have two or more time coordinates. 
To go from $p$ to $p'$, one simply forgets the directions. 
To go to $p''$, one looks for almost-rooted trees compatible with $p'$, with the additional restriction that the number of outgoing edges must be odd for every (non-root) vertex. 
This restriction arises from the fact that each commutator in the Magnus recursion produces the anti-symmetric fuzzy Wightman function $W_A^{\#}$ in \eqref{bw-fuzzy-A} which is an odd function of $W_A$. 

Each $p''$ comes with further multiplicative factors. 
One is the sign factor; compare  $p''$ to $p$ and count how many edges need to be flipped. 
Unlike in the color case, the edges can be flipped individually without restriction; 
the sign flip concerns time ordering and not operator ordering. 
The sign factor is called $(-1)^{\ell(p'')}$. The other counts powers of $(-1/4)$ originating from the fuzzy Wightman function in \eqref{bw-fuzzy-A}. A vertex $v$ in $p''$ with $2k_v+1$ outgoing edges contributes $(-1/4)^{k_v}$. If $m(p'')$ is the sum of all $k_v$, the net factor becomes $(-1/4)^{m(p'')}$.

\begin{figure}[htbp]
    \centering
    \includegraphics[width=0.75\linewidth]{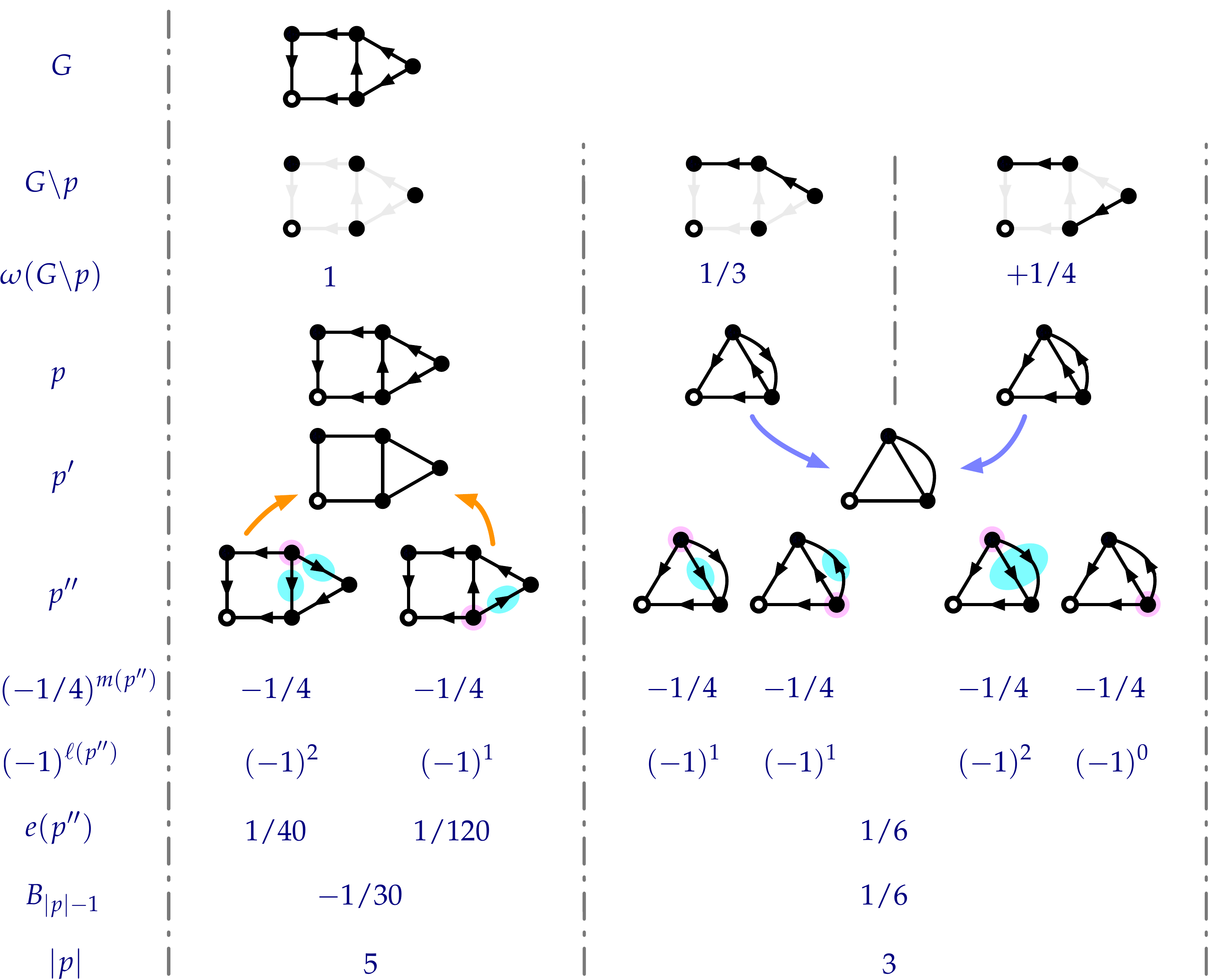}
    \caption{Murua formula in the BW basis for a pentagon graph.}
    \label{fig:MS-penta-bw}
\end{figure}

An example is given in Fig.~\ref{fig:MS-penta-bw}. 
The light blue shades in $p''$ count $\ell(p'')$. 
The pink shades indicate vertices contributing to $(-1/4)^{m(p'')}$. 
The net result of the formula can be summarized as 
\begin{align}
\begin{split}
    \omega(G)
    \,&=\,
\lrp{
\begin{aligned}[c]
    &
        \left(-\frac{1}{4}\right) \cdot 1\cdot  \left( \frac{(-1)^2}{40} + \frac{(-1)^1}{120} \right) \left(-\frac{1}{30}\right)\cdot 5 
    \\
    &
        + \left(-\frac{1}{4}\right) 
        \bb{
            \frac{1}{3} \cdot \frac{(-1)^1+(-1)^1}{6}  
            +  \frac{1}{4} \cdot \frac{(-1)^2+(-1)^0}{6} 
        }
        \cdot \frac{1}{6} \cdot{3}
\end{aligned}
}   
    \,=\,
        \frac{1}{240} 
    \,. 
\end{split}
\end{align}

\paragraph{Contraction rules}

The edge-contraction rules are also available in the BW basis.  
The first rule concerns the cut propagator: 
\begin{align}
\begin{split}
    \omega(G_{[1\textcolor{gray}{\cdots}2]}) 
    \,=\,
        \omega(G_{[1 \vert 2]\text{conn}} )
    \,.
\end{split}
\label{bw-contraction-cut}
\end{align} 
The value of $\omega$ remains the same when the cut propagator is ``erased,'' as long as the graph remains connected. 
This phenomenon is implied by the fuzzy propagator discussed in \Sec{sec:propagators}. 
It generalizes the observation of \rcite{Pichini} (up to convention-dependent powers of the factor 2). 
The triangle example in Fig.~\ref{fig:trans-triangle} and the quadrangle example in Fig.~\ref{fig:bw-1-loop-quad} illustrate how the cut rule in \eqref{bw-contraction-cut} works. 

The second rule concerns the retarded propagator, 
\begin{align}
\begin{split}
    \omega(G_{[1\leftarrow 2]}) 
    \mem+\,
    \omega(G_{[1 \rightarrow 2]})
    \,=\,
    - \omega(G_{[1 \cdot 2]})   
    \,.
\end{split}
\label{bw-contraction-retarded}
\end{align}
The proof of \eqref{bw-contraction-retarded} is similar to that of \eqref{contraction-color-a}. When the edge [1-2] belongs to the partition $p$, the two graphs of the LHS give the same $p''$ with $\Delta\ell=1$, which cancel each other. Thus, the remaining terms on the LHS arise when the edge [1-2] belongs to $G\backslash p$, which perfectly matches the RHS by recursively applying \eqref{bw-contraction-retarded} to the $\omega$ in $G\backslash p$.

\begin{figure}[htbp]
    \centering
    \includegraphics[width=0.9\linewidth]{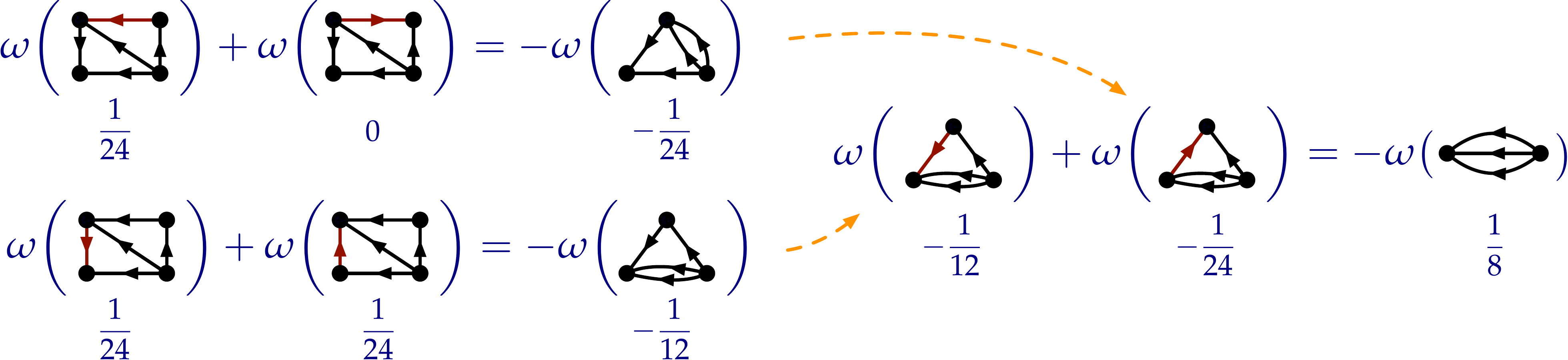}
    \caption{How the BW contraction rule works.} 
    \label{fig:contraction-bw}
\end{figure}

Fig.~\ref{fig:contraction-bw} shows some examples of the contraction rule. 
After dispensing with cut propagators using \eqref{bw-contraction-cut}, we are left with loop graphs composed entirely of retarded propagators. The number of loops, $L$, is even because of color parity.
$L$ is the same between the two sides of the contraction rule. 
So, unlike in the color basis, we cannot run a recursion relating $L$-loop graphs to $(L-1)$-loop graphs. 
But the contraction rule reduces the number of edges by 1. 
Applying the rule repeatedly, we can reduce all primary loops (triangle, quadrangle, etc.) to 
banana loops. 
The edges of the banana loops can be removed in pairs, connecting $L$-loop graphs to $(L-2)$-loop graphs. 

For a fixed graph topology (undirected graph as a family of oriented graphs), knowing the $\omega$ value of one graph 
is sufficient to determine $\omega$ of all other graphs in the same topology. 
Unlike in the color basis, we do not have graphs with $\omega=0$. 
We can run the Murua formula for any one representative of the topology and use the contraction rule 
to determine $\omega$ of all the other graphs. 
When flipping the direction of edges to apply the contraction rule, we have to avoid cyclic graphs. 
It is easy to show that all graphs in the same \emph{primary} loop topology can be connected by flipping directions 
without ever encountering a cyclic graph along the way.

\paragraph{Effective field theory matching}

In QFT, it is a common practice to begin with an ultraviolet (UV) theory 
with heavy and light fields, and integrate out heavy fields to obtain an infrared (IR) theory with light fields only. 
A full-fledged discussion of renormalization is beyond the scope of this paper, but 
let us examine the role of the contraction rule in the process of integrating out heavy fields,  
in a similar vein to the discussions of \rcite{KKKL}. 

\begin{figure}[htbp]
    \centering
    \includegraphics[width=0.5\linewidth]{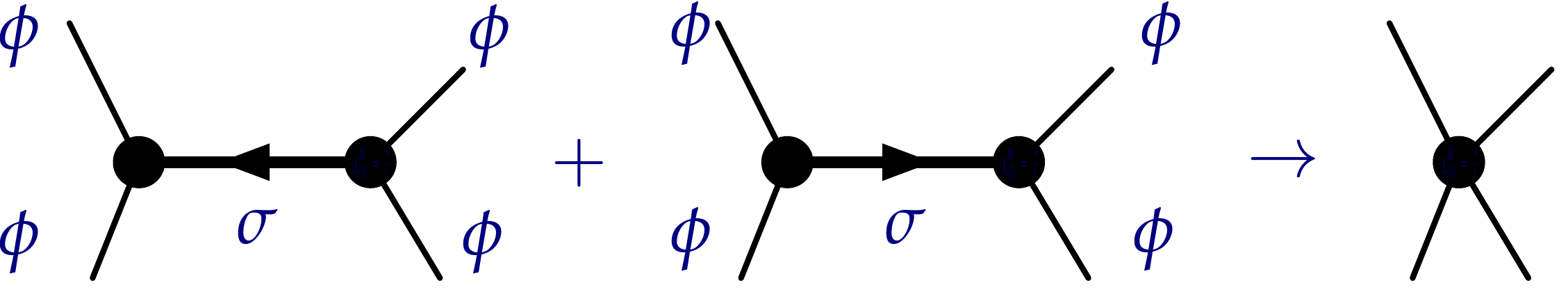}
    \caption{Integrating out a heavy field from a Magnusian.}
    \label{fig:contraction-EFT}
\end{figure}

Consider a toy model with the UV Lagrangian, 
\begin{align}
    \mathcal{L}_\text{UV} 
    \,=\,
    -\frac{1}{2}(\partial\phi)^2  - \frac{1}{2}m^2\phi^2 - \frac{\lambda_0}{4!} \phi^4 -  \frac{1}{2}(\partial\sigma)^2 - \frac{1}{2}M^2\sigma^2 + \frac{1}{2}\mu \sigma \phi^2 \,. 
\end{align}
Integrating out the heavy field $\sigma$, we obtain the IR Lagrangian, 
\begin{align}
\label{EFT-IR}
    \mathcal{L}_\text{IR} 
    \,=\,
    -\frac{1}{2}(\partial\phi)^2  - \frac{1}{2}m^2\phi^2- \frac{\lambda_1}{4!} \phi^4  \,,
    \qquad \frac{\lambda_1}{4!} \,=\, \frac{\lambda_0}{4!}  -\frac{\mu^2}{8M^2} \,. 
\end{align}
Suppose we compute the quantum Magnusian of the UV theory and take the $M\rightarrow \infty$ limit. 
The result should match the Magnusian of the IR theory, up to terms suppressed at least by $m^2/M^2$. 
The BW contraction rule for retarded propagators in \eqref{bw-contraction-retarded} 
ensures that the $\omega$ values of the two diagrams of the UV theory add up to give the $\omega$ value of the IR theory. 
Fig.~\ref{fig:contraction-EFT} shows one of the three channels through which the $\sigma$ field mediates 
the interaction of $\phi$ fields. The effective quartic coupling constant of the IR theory is 
\begin{align}
    \lambda_0 - 3\times \mu \frac{1}{p^2 + M^2} \mu  \;\; \approx \;\; \lambda_0 - 3 \frac{\mu^2}{M^2} \,, 
\end{align}
in agreement with the expectation from the Lagrangian in \eqref{EFT-IR}.

Those UV diagrams with a heavy particle sitting on the cut propagator do not affect the matching. 
The cut propagator would enforce the mass-shell condition of the heavy particle, 
but as far as the matrix elements of light fields are concerned, 
the heavy mass-shell condition is out of reach. 
The Magnusian of the IR theory does not admit any diagram corresponding to heavy cut diagrams of the UV theory. 

% \newpage
\subsection{Change of Basis} 
\label{sec:color-vs-BW-redux}

Recall that the map between the color basis and the BW basis is based on taking linear combinations of (fuzzy) propagators. 
Not surprisingly, the $\omega$ values of the two bases satisfy some linear relations.

\paragraph{From color to BW}

Having seen how the two bases are related to each other in examples, we now discuss closed form formulas that connect the $\omega$ values in one basis to that of the other basis. 
We begin with the map from the color basis to the BW basis.

\begin{figure}[htbp]
    \centering
    \includegraphics[width=0.5\linewidth]{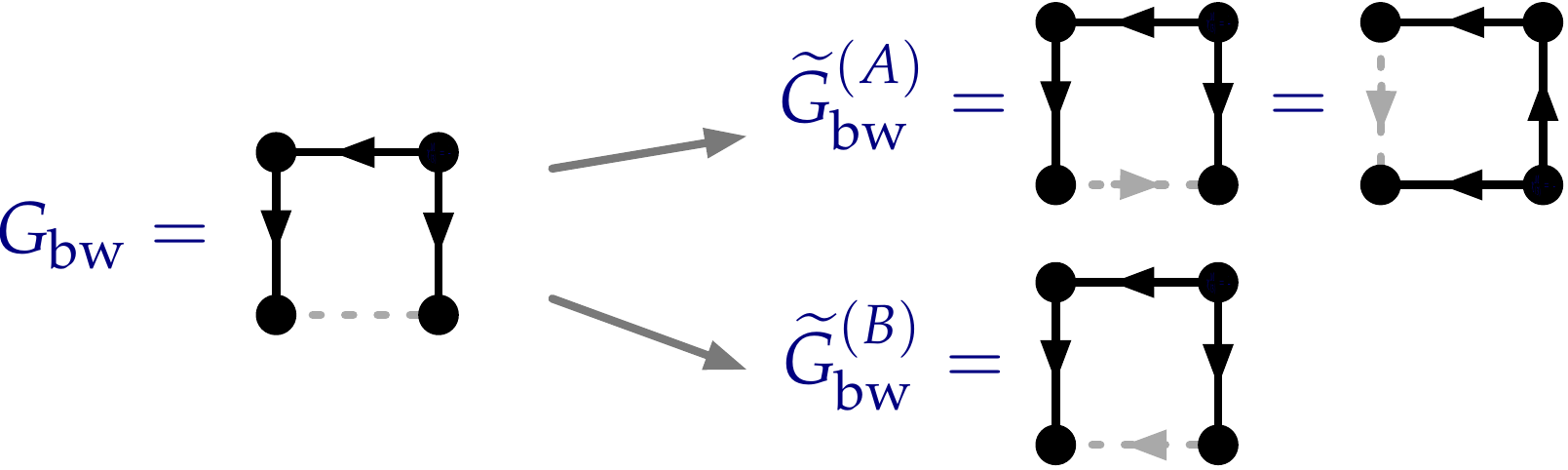}
    \caption{Two possibilities for directed completion.}
    \label{fig:bw-dir-choice}
\end{figure}

In the color basis, all edges are directed. In the BW basis, the retarded propagators are directed, but the cut propagators are not. Given a graph in the BW basis, $G_\text{bw}$, the first step is to introduce the notion of ``directed completion"; we assign directions to the cut propagators in a way that does not create any cycle. 
In physics terms, we assign a time ordering of the cut propagators that is compatible with the pre-existing time ordering of the retarded propagators. 
In general, the directed completion is not unique, so we can make a choice. 
See Fig.~\ref{fig:bw-dir-choice} for an example of $G_\text{bw}$ which admits two non-equivalent completions. 

Let $\widetilde{G}_\text{bw}$ be any one of the directed completions of $G_\text{bw}$. 
Define $X(\widetilde{G}_\text{bw})$ to be the set of all colored graphs $G_\text{color}$ which coincide with $\widetilde{G}_\text{bw}$ 
when ``bleached" (rid of the color). 
To compute $\omega_\text{bw}(G_\text{bw})$, we take the following weighted sum,  
\begin{align}
    (-1)^{(L-E_\text{cut})/2}\,
        \omega_\text{bw} (G_\text{bw}) 
    \,=\,
        \frac{1}{2^{\text{ret}(G_\text{bw})}} \sum_{G_\text{color}\in X(\widetilde{G}_\text{bw})}
        (-1)^{\text{ret-blue}}\,
            \omega_\text{color}(G_\text{color}) 
    \,. 
    \label{omega-color-to-bw}
\end{align}
Here, $\text{ret}(G_\text{bw})$ is the number of retarded propagators of $G_\text{bw}$. 
The sign factor $(-1)^\text{ret-blue}$ depends on the number of blue edges of $G_\text{color}$ that match the retarded propagators of $\widetilde{G}_\text{bw}$. 
An important fact is that the resulting $\omega_\text{bw}$ is independent of the choice of $\widetilde{G}_\text{bw}$. 

Physically,  $\omega_\text{bw}$ being independent of the choice is a manifestation of Lorentz invariance. 
If the directed completion of a cut propagator is not unique, there always exists a choice of spacetime points consistent with the time-orderings implied by the causal propagators where the relative time-ordering of vertices connected by the considered cut propagator can be reversed by a Lorentz transform. 

Let us apply the formula to the example in Fig.~\ref{fig:bw-dir-choice}. 
Working with $\tilde{G}^{(A)}_\text{bw}$, we import the $\omega$ values of the colored trapezoids from Fig.~\ref{fig:bicolor-trapezoid}. The formula gives
\begin{align}
    \omega_\text{bw}  
    \,=\,
        \frac{1}{2^3} \cdot 2\cdot 
        \bb{
            - \frac{1}{4} + \frac{1}{12} + \frac{1}{12} - \frac{1}{12} + 0 - \frac{1}{6} -\frac{1}{6} + \frac{1}{6}
        }
    \,=\,
        -\frac{1}{12} 
    \,.
\end{align}
Here, $1/2^3$ is from the $1/2^{\text{ret}}$ factor. The other factor 2 in front is due to the color parity doubling. The eight numbers in the parenthesis come from the sum over the eight graphs in Fig.~\ref{fig:bicolor-trapezoid}, modulated by $(-1)^{\text{ret-blue}}$. 

If we choose to work with $\tilde{G}^{(B)}_\text{bw}$ instead, we import the $\omega$ values of the colored diamonds from Fig.~\ref{fig:bicolor-diamond}. The formula then gives 
\begin{align}
    \omega_\text{bw} 
    \,=\,
        \frac{1}{2^3} \cdot 2\cdot \bb{
        -\frac{1}{6} + 0 + 0 + 0 - \frac{1}{12} -\frac{1}{12} - \frac{1}{12} + \frac{1}{12}
    } 
    \,=\,
        -\frac{1}{12} 
    \,.
\end{align}
We confirm that $\omega_\text{bw}$ is independent of the choice between $\tilde{G}^{(A)}_\text{bw}$ and $\tilde{G}^{(B)}_\text{bw}$.

\paragraph{From BW to color} 

The inverse formula of \eqref{omega-color-to-bw} reads 
\begin{align}
    \omega_\text{color} (G_\text{color}) 
    \,= 
        \sum_{G_\text{bw} \in Y(G_\text{color})}   
        \frac{(-1)^{\text{blue-ret}}}{2^{\text{cut}(G_\text{bw})}}\,
        \BB{
            (-1)^{(L-E_\text{cut})/2} \omega_\text{bw}(G_\text{bw}) 
        }
    \,.
    \label{omega-bw-to-color}
\end{align}
Here, the colored graph $G_\text{color}$ is the input. We define $Y(G_\text{color})$ to be the set of all BW graphs whose suitably directed $\tilde{G}_\text{bw}$ agrees with the bleached $G_\text{color}$. 

\begin{figure}[htbp]
    \centering
    \includegraphics[width=0.64\linewidth]{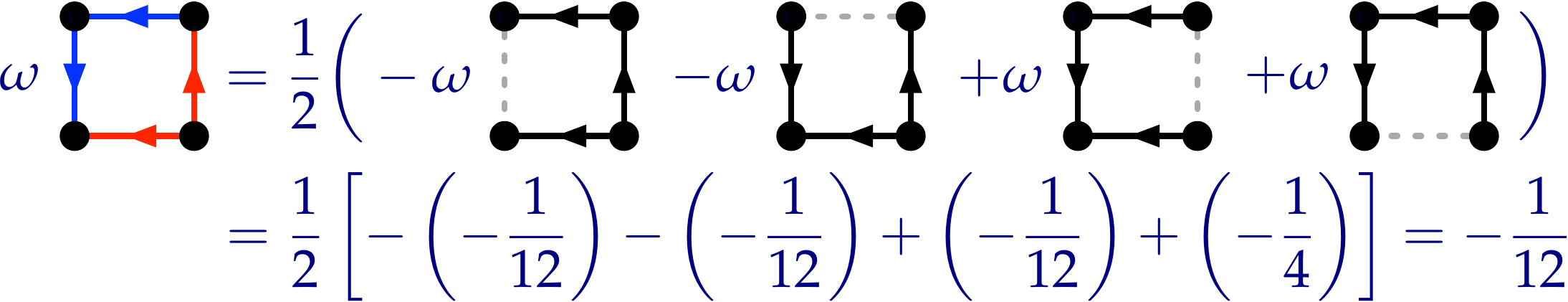}
    \caption{Computing $\omega$ in the color basis from those of the BW basis.}
    \label{fig:color-from-bw}
\end{figure}

An example is given in Fig.~\ref{fig:color-from-bw}. 

It is straightforward to show that \eqref{omega-color-to-bw} and  \eqref{omega-bw-to-color} are mutually inverse functions. Inserting \eqref{omega-color-to-bw} into \eqref{omega-bw-to-color}, we have 
\begin{align}
\begin{split}
    \omega_\text{c} (G_\text{c}) 
    \,&=
        \sum_{G_\text{bw} \in Y(G_\text{c})}   
            \frac{(-1)^{\text{blue-ret}}}{2^{\text{cut}(G_\text{bw})}} 
            \frac{1}{2^{\text{ret}(G_\text{bw})}} 
        \sum_{G'_\text{color}\in X(\widetilde{G}_\text{bw})}
            (-1)^{\text{ret-blue}'}
            \omega_\text{c}(G'_\text{c})
    \\
    \,&=\, 
        \frac{1}{2^{E(G_\text{c})}}\mem
        \sum_{Y,X}\,
            (-1)^\text{blue-ret}\mem (-1)^{\text{ret-blue}'}\,
            \omega_c(G_\text{c}')
    \,. 
\end{split}
\end{align}
The sign factors are all $(+1)$ for $G'_\text{c} = G_\text{c}$. 
For any other $G'_\text{c}$, there are equal numbers of $(+1)$ and $(-1)$ summing up to zero. 

The two-way transformation rules, \eqref{omega-color-to-bw} and \eqref{omega-bw-to-color}, map the contraction rule on one side to the contraction rule on the other side, providing a further consistency check.

%\newpage 
\paragraph{Final example: 4-point, 3-loop}

In the BW basis, the number of cut propagators, $E_\text{cut}$, is bounded by the number of loops $L$ as $0 \le E_\text{cut} \le L$.  
The color parity as stated in \eqref{color-parity-BW} dictates that $L-E_\text{cut} \in 2\mathbb{Z}$. 
The case $L=E_\text{cut}$ is called ``maximally cut.'' 
For an example with non-maximally cut graphs, we consider 
the 4-point, 3-loop graphs. 

\begin{figure}[htbp]
    \centering
    \includegraphics[width=0.95\linewidth]{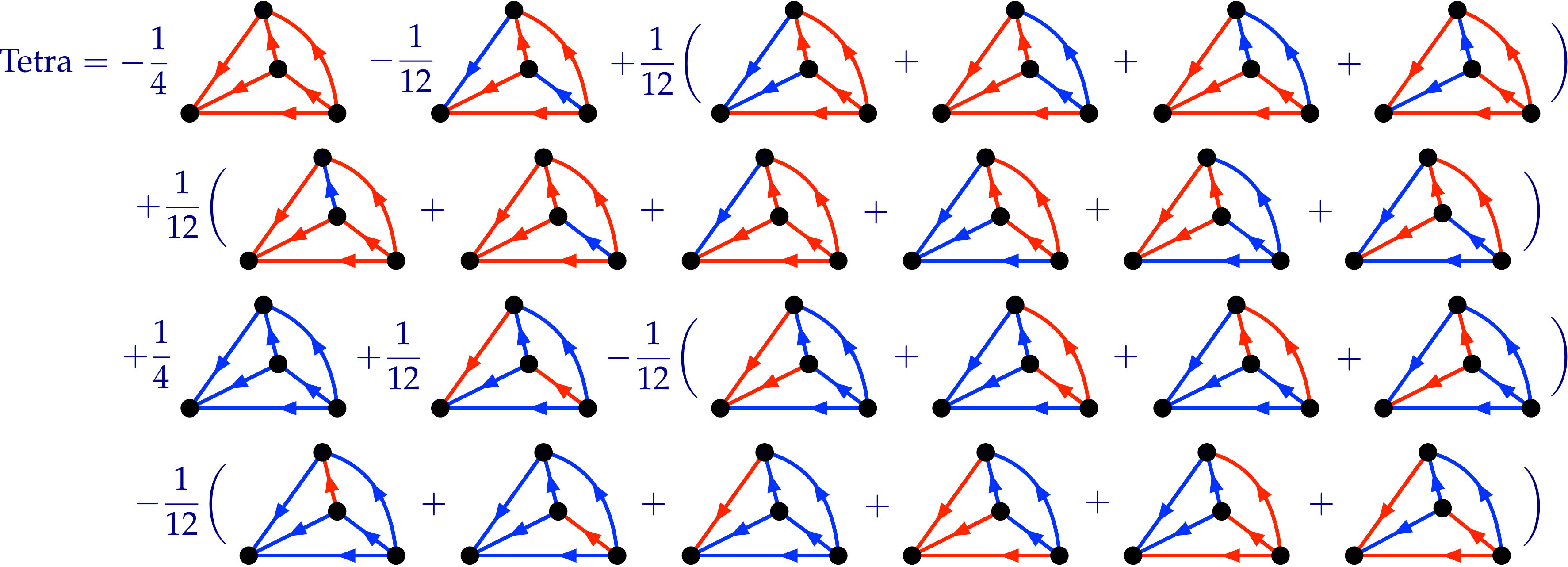}
    \caption{4-point, 3-loop graphs in the color basis. Graphs with $\omega=0$ are omitted to save space.}
    \label{fig:4pt-3loop-color}
\end{figure}

Fig.~\ref{fig:4pt-3loop-color} shows the result in the color basis. It shows 24 out of 64 colored graphs that have non-zero $\omega$ values. The first two lines and the last two lines are color-flip images of each other.

\begin{figure}[htbp]
    \centering
    \includegraphics[width=0.85\linewidth]{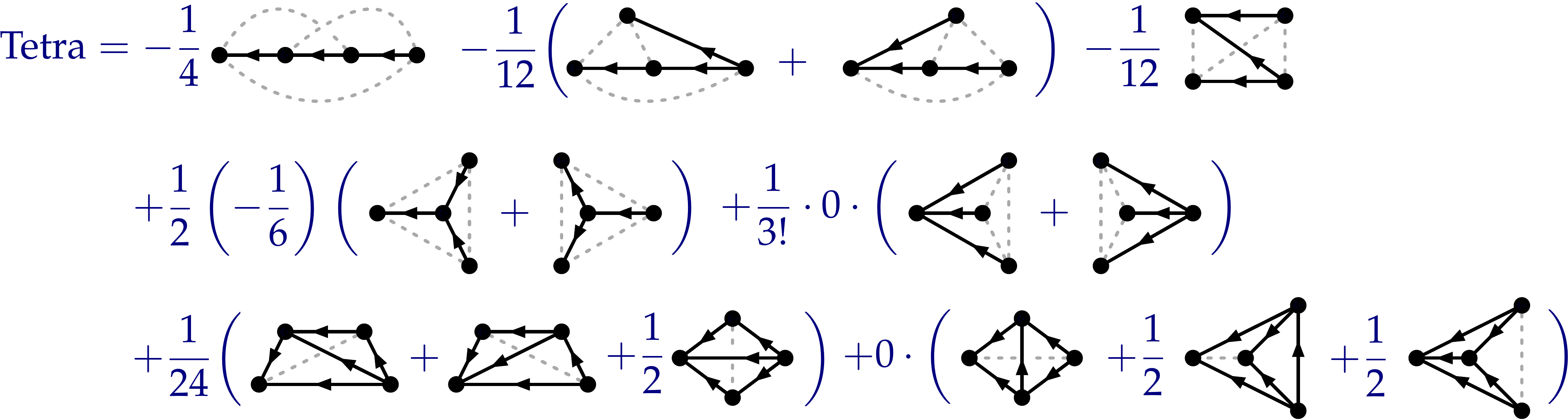}
    \caption{4-point, 3-loop graphs in the BW basis.
    }
    \label{fig:4pt-3loop-bw} 
\end{figure}

Fig.~\ref{fig:4pt-3loop-bw} shows the same result in the BW basis. 
The colored version has no symmetry factor to consider. When it is translated to the BW basis, 
the symmetry factors should be taken into account to extract the $\omega$ values.

%%%%%%%%%%%%%%%%%%%%%%%%%%%%%%%%%%%%%%%%%%%%%%%%%%%%%%
\section{Diagrammar at a Glance}
\label{sec:summary-diagrammar}

The discussions in \Secs{sec:Q-Mag}{sec:Murua} establish a purely diagrammatic formulation of the quantum Magnusian
and the Murua coefficient.
To make this point explicit and clear,
we provide a compact summary of
the quantum Magnusian diagrammar
as the core message of this paper.

\paragraph{Sum over graphs}

The quantum Magnusian is given by a sum over graphs. 
We have two equivalent expressions, one in the color basis and the other in the BW basis. 
\begin{align}
\begin{split}
\label{Mag-all}
    {\EikonalSignC}\chi 
    \,&=\,
        \sum_{G_c}\,
            \frac{\omega_c(G_c)}{\sigma_c(G_c)}
            \,
            \hbar^{L(G_c)}
            \,
            i^{V(G_c)-1}
            \,
            \mathcal{I}(G_c) 
    \\
    \,&=\,
        \sum_{G_\text{bw}}\,
            \frac{\omega_\text{bw}(G_\text{bw})}{\sigma_\text{bw}(G_\text{bw})}
            \,
            \hbar^{L(G_\text{bw})}
            \,
            \mathcal{I}(G_\text{bw}) 
    \,.
\end{split}
\end{align}
The $\omega$ values of descendant graphs are completely determined by those of primary graphs.
The graph integral is given by 
\begin{align}
\label{IG-OG-times-PG-re}
    \mathcal{I}(G) 
    \,=\,
        \int \prod_{i=1}^{|G|} d^dx_i\,\,
            \mathcal{O}_G(x_1, \cdots , x_{|G|})
            \,
            \mathcal{P}_G(x_1, \cdots , x_{|G|})
    \,.
\end{align}
The $\mathcal{P}_G$ factor denotes the product of 
all the propagators appearing in the graph, 
and $\mathcal{O}_G$ is the product of all vertex factors. 

\paragraph{Primary graphs} 
The time ordering in the Magnus expansion is more intuitive in position space, but the propagators are more succinctly described in momentum space. %
Our convention for a momentum space two-point function is 
\begin{align}
    G (k) 
    \,=
        \int d^dx\,\,
            e^{-ik \cdot x}\, G (x;0) 
    \,,
\end{align}
which means $G(k)$ should be analytic on the upper half plane of $k^0$ if $G(x;0)$ vanishes for $t < 0$.
The colored propagators read 
\begin{align}
\begin{split}
    G_\text{\textcolor{red}{red}}(k) 
    \,=\,
        \frac{i}{2E_{\vec{k}}\bigbig{
            k^0 - E_{\vec{k}} + i 0^+
        }}
    \,,\quad
    G_\text{\textcolor{blue}{blue}}(k)
    \,=\,
        \frac{i}{2E_{\vec{k}}\bigbig{
            k^0 + E_{\vec{k}} + i 0^+
        }}
    \,, 
\end{split}
\end{align}
with $E_{\vec{k}} = (\vec{k}^2 + m^2)^{1/2}$.
The Hermiticity is realized as $[G_\text{\textcolor{red}{red}}(k)]^* =  G_\text{\textcolor{blue}{blue}}(-k)$.
The retarded and cut propagators in the BW basis are
\begin{align}
\begin{split}
    G_\text{ret}(k) 
    \,&=\,
        \frac{1}{k^2 +m^2 -i0^+ \text{sgn}(k^0)} = \frac{1}{ -\bigbig{k^0 + i 0^+}^{\nem2} + \vec{k}^2 + m^2 }
    \,,\\
    G_\text{cut}(k)
    \,&=\,
        \pi\, \delta(k^2+m^2)
    \,.
\end{split}
\end{align}
The vertex factors are the same in the two bases. 
A valence $m$ vertex reads 
\begin{align}
    \mathcal{V}_m 
    \,=\,
        \frac{\partial^m \mathcal{V}}{\partial \phi^m} 
    \,.
\end{align}

\paragraph{Descendant graphs} 

In the color basis, any red or blue propagator can turn into a banana loop of the same color with multiplicity $r$, producing a multi-propagator weighted by a symmetry factor:
\begin{align}
  \frac{1}{r!}\, (G_\text{color})^r \,.
\end{align} 
In the BW basis, a banana loop with $r$ retarded and $s$ cut propagators is weighted by 
\begin{align}
   \left(-\frac{1}{4}\right)^{\nem\nem\lfloor r/2 \rfloor} \frac{1}{r! s!}\, (G_\text{ret})^r \mem(G_\text{cut})^s \,.
\end{align}
Odd/even values of $r$ correspond to the fuzzy retarded/cut propagators, 
respectively. 

% \newpage 
\paragraph{Edge-contraction rules}

The edge-contraction rule in the color basis is
\begin{subequations}
\label{contraction-color-re}
\begin{align}
    \omega(G_{\textcolor{red}{[1\leftarrow 2]}}) -\omega(G_{\textcolor{blue}{[1 \rightarrow 2]}}) \,&=\, -\omega(G_{[1\cdot 2]}) \,,
    \label{contraction-color-a-re}
    \\
    \omega(G_{\textcolor{red}{[1\leftarrow 2]}}) +\omega(G_{\textcolor{blue}{[1 \leftarrow 2]}}) \,&=\, \omega(G_{[1 \vert 2]\text{conn}}) \,. 
    \label{contraction-color-b-re}
\end{align}
\end{subequations}
The edge-contraction rule in the BW basis is
\begin{subequations}
\label{bw-contraction-re}
    \begin{align}
    \omega(G_{[1\textcolor{gray}{\cdots}2]}) 
    \,&=\, 
        \omega(G_{[1 \vert 2]\text{conn}} )
    \,, 
    \label{bw-contraction-cut-re}
    \\
    \omega(G_{[1\leftarrow 2]}) + \omega(G_{[1 \rightarrow 2]})
    \,&=\,
        - \omega(G_{[1 \cdot 2]})   
    \,.
\label{bw-contraction-retarded-re}
\end{align}
\end{subequations}
Here, $G_{[1 \cdot 2]}$ is the graph obtained from $G_{[1 \rightarrow 2]}$ or $G_{[1\leftarrow 2]}$ by contracting the edge to a vertex, and $G_{[1 \vert 2]\text{conn}}$ is the connected graph obtained from $G_{[1 \rightarrow 2]}$ or $G_{[1\leftarrow 2]}$ by removing the edge; otherwise, it vanishes. See Fig.~\ref{fig:contraction-monochrome} for examples.

\paragraph{Basis change rules} 

To map the $\omega$ values from the color basis to the BW basis or vice versa, we can use the two formulas below: 
\begin{subequations}
\label{basis-change-re}
    \begin{align}
    (-1)^{(L-E_\text{cut})/2}\,
        \omega_\text{bw} (G_\text{bw})
    \,&=\,
        \frac{1}{2^{\text{ret}(G_\text{bw})}}
        \sum_{G_\text{color}\in X(\widetilde{G}_\text{bw})} (-1)^{\text{ret-blue}} \omega_\text{color}(G_\text{color}) 
    \,, 
    \label{omega-color-to-bw-re}
    \\
    \omega_\text{color} (G_\text{color}) 
    \,&=
        \sum_{G_\text{bw} \in Y(G_\text{color})}    \frac{(-1)^{\text{blue-ret}}}{2^{\text{cut}(G_\text{bw})}}\,
        \BB{
            (-1)^{(L-E_\text{cut})/2}\, \omega_\text{bw}(G_\text{bw})
        }
    \,.
    \label{omega-bw-to-color-re}
\end{align} 
\end{subequations}

\paragraph{Use}
To connect with the practical needs of physicists,
recall \eqref{impulse}.
Once the quantum Magnusian is computed,
the impulse of quantum observables
arises via the formula: 
\begin{align}
%% This equation is written in the New Convention
    \expval{\mathcal{O}}_{\hnem\text{out}}
    \,=\,
        \expval{
            e^{{{\MinusEikonalSignC}\nem\ad_{\chi/i\hbar}}}\mem
            \mathcal{O}
        }_{\text{in}}
    \,.
\end{align}
% which admits a smooth classical limit as well.

%\newpage
\section{Conclusion} \label{sec:conclusion}

In this paper,
we completed the diagrammar of the quantum Magnusian
by combining and extending
insights from pioneering works \cite{Pichini,PSFOR-S}.
This diagrammar is summarized in \Sec{sec:summary-diagrammar}.
Although we presumed the specific context of relativistic scalar QFTs
for concreteness,
it should be emphasized that the Magnusian is a much more general and broader concept
as remarked in \Sec{sec:review}.
Hence, up to mild variations,
we expect the relevance of our diagrammar
to generic QFTs,
to other branches of physics,
and even beyond physics.

On top of the first-principles frameworks
\cite{Pichini,PSFOR-S},
the purely diagrammatic approach pursued in this paper
facilitates an efficient algorithm for constructing the quantum Magnusian,
based on the edge-contraction rules.

Although we now have sufficiently efficient methods for
computing the quantum Magnusian,
our understanding of its physics is far from complete.
Below, we list a few questions regarding the physics of the quantum Magnusian.

\begin{itemize}
\item 
    The modern understanding of QFTs is that they should be understood as effective descriptions that require regularization and renormalization to make sensible predictions.
    How can we prove renormalizability of the quantum Magnusian without resorting to the finiteness of the $S$-matrix?
\item 
    The singularities of the $S$-matrix elements are known to be associated with particle production, which can be explored through the Landau equations~\cite{Landau:1959fi}.
    What is the physics behind the singularities (if any) of the quantum Magnusian, and what are the analogues of the Landau equations for exploring its singular loci?
\item 
    Many computations in gauge theories are organized in the $1/N$ expansion, as in the planar sector of $\mathcal{N} = 4$ SYM.
    Is there any qualitative difference (e.g., organization of color factors) between the $1/N$ expansion of $S$-matrix elements and the $1/N$ expansion of Magnus amplitudes? 
\item 
    And lastly, Wilson lines are computed as \emph{path-ordered} exponentiated operators.
    How can we repurpose the mathematical machinery for diagrammatic expansion of the quantum Magnusian for computations involving Wilson lines?
\end{itemize}

Another set of questions we can ask about the quantum Magnusian concerns whether it is compatible with more modern methods developed for computing scattering amplitudes, which avoid explicit references to vertex rules and Feynman diagrams. 
Examples of such methods include generalized unitarity~\cite{Bern:1994zx,Bern:1994cg} and the color-kinematics dual form of gauge theory integrands for gravity computations~\cite{Bern:2008qj,Bern:2010ue}, 
which relies on the fact that a family of Feynman diagrams that shares a common subgraph can be grouped together.
For tree diagrams, \rcite{KKKL} identified a sum rule dubbed ``Feynman reduction'' as a possible mechanism for applying such modern tools to the computation of the (classical) Magnusian. 
It would be interesting to explore whether similar sum rules exist for diagrams with loops, 
and how such modern methods can be applied to the computation of the quantum Magnusian.

In \rcite{KKKL}, the Hopf algebra behind the Magnus expansion, introduced by Calaque, Ebrahimi-Fard, and Manchon~\cite{Calaque_2011} (see also \cite{Chartier_2010}), 
was extended from rooted trees to non-rooted trees, providing extra conceptual and technical layers to 
the tree diagrammar of the Magnusian. A natural question is how to extend the Hopf algebra further to loop graphs 
endowed with the color/BW basis of this paper. 
This very question will be the main topic of a companion paper~\cite{our-math-paper}. 

\vskip 1cm 

\acknowledgments 

JWK would like to thank Andreas Brandhuber, Paolo Pichini, and Gabriele Travaglini for helpful discussions. 
Three of the authors (JWK, SL, JL)  would like to thank the Erwin Schr\"odinger International Institute for Mathematics and Physics (ESI), University of Vienna (Austria), for the opportunity to participate in the Thematic Programme “Amplitudes and Algebraic Geometry" in 2026, where a significant part of this work was accomplished, and for the support given.
JHK is supported by the Department of Energy (Grant No.~DE-SC0011632) and by the Walter Burke Institute for Theoretical Physics. The work of SK and SL is supported by the National Research Foundation of Korea (NRF) grant NRF RS-2024-00351197 and KIAS grant PG006002. The work of JL is supported by the Austrian Science Fund (FWF), PAT 9039323, Grant-DOI 10.55776/PAT9039323.

\newpage 

\bibliography{phys-references}

\end{document}